\def\year{2015}
\documentclass[letterpaper]{article}
\usepackage{aaai}
\usepackage{times}
\usepackage{helvet}
\usepackage{courier}

\usepackage{epsf}
\usepackage{epsfig}
\usepackage{mathrsfs}
\usepackage{amsmath}
\usepackage{caption}
\DeclareCaptionType{copyrightbox}
\usepackage{subfig}
\usepackage{graphics}
\usepackage{balance}
\usepackage{graphicx}
\usepackage{balance}
\usepackage[boxed]{algorithm2e}
\usepackage{epstopdf}
\usepackage{slashbox}
\usepackage{url}
\usepackage{multirow}
\usepackage{hhline}
\usepackage{pifont}
\usepackage{array,ragged2e}
\usepackage{lscape}

\usepackage{amsfonts}

\usepackage{pifont}

\usepackage{xcolor}

\newcommand{\tbl}{\caption}

\newcommand{\increaseevent}{\ensuremath{I}}
\newcommand{\specialevent}{\ensuremath{S}}
\newcommand{\externalitiesevent}{\ensuremath{E}}

\newcommand{\venue}{\ensuremath{v}}

\newcommand{\checkinst}{\ensuremath{c_{a}}}

\newcommand{\norcheckinst}{\ensuremath{c}}

\newcommand{\uniquevt}{\ensuremath{p_{a}}}

\newcommand{\median}{\ensuremath{m}}

\newcommand{\types}{\ensuremath{{\cal T}}}
\newcommand{\type}{\ensuremath{T}}

\newcommand{\loyalty}{\ensuremath{{\tt \lambda}}}

\newcommand{\likes}{{\tt \iota}}

\newcommand{\tips}{\ensuremath{{\tt N_t}}}
\newcommand{\duration}{{\tt D}}
\newcommand{\specialCount}{{\tt N_s}}

\newcommand{\neighborhood}{\mathcal{N}}
\newcommand{\neighborhooddensity}{\mathcal{\rho}}
\newcommand{\areapopularity}{\mathcal{\phi}}
\newcommand{\competitiveness}{\mathcal{\kappa}}
\newcommand{\entropy}{\mathcal{\varepsilon}}
\newcommand{\specialTypeVector}{\boldsymbol\xi_s}

\newcommand{\boots}{\ensuremath{\mathcal{B}}}

\begin{document}
%
\title{Analyzing and Modeling Special Offer Campaigns in\\ Location-based Social Networks}
\author{Ke Zhang \\
University of Pittsburgh \\
kez11@pitt.edu\\
\And
Konstantinos Pelechrinis \\
University of Pittsburgh \\
kpele@pitt.edu \\
\And
Theodoros Lappas \\
Stevens Insitute of Technology \\
tlappas@stevens.edu
}
\maketitle
\begin{abstract}
\begin{quote}
The proliferation of mobile handheld devices in combination with the technological advancements in mobile computing has led to a number of innovative services that make use of the location information 
available on such devices.
Traditional yellow pages websites have now moved to mobile platforms, giving the opportunity to local businesses and potential, near-by, customers to connect.
These platforms can offer an affordable advertisement channel to local businesses.
One of the mechanisms offered by location-based social networks (LBSNs) allows businesses to provide special offers to their customers that connect through the platform.
We collect 
a large time-series dataset from approximately 14 million venues on Foursquare 
and analyze the performance of such campaigns
using randomization techniques and (non-parametric) hypothesis testing with statistical bootstrapping.
Our main finding indicates that this type of promotions are not as effective as anecdote success stories might suggest.
Finally, we design classifiers by extracting three different types of features that are able to provide an educated decision on whether a special offer campaign for a local business will succeed or not both in short and long term.
\end{quote}
\end{abstract}

\section{Introduction}
\label{sec:intro}

During the last years a number of location-based services and social media has emerged mainly due to the rapid proliferation of mobile handheld devices in combination with the technological advancements in mobile computing.
People can use these devices to obtain a wide range of information related to the geographic area they are currently in.
Web services that have traditionally aimed at digitally connecting people with local businesses (e.g., Yelp, Urbanspoon etc.) are transforming to mobile.
This transformation facilitates a real-time interaction between the involved parties through a two-way communication channel.
For instance, Yelp users can initiate a mobile application on their devices and get instant information for locales that are within their reach.  

However, this mobile transformation of ``yellow pages'' services is beneficial to local businesses as well.
Not only can they be discovered  by people that are near-by, but most importantly they have an immediate way of advertising to potential customers.
One of the advertisement mechanisms allows venues to use such mobile platforms to provide special offers to customers that connect with them through these services.
For instance, a venue on Foursquare can offer special deals to people that check-in to the locale through the application. 
The same is true for Yelp users, 
even though the actual details might differ.
This can potentially be an inexpensive way of advertisement for local businesses to people that are nearby and actually have the potential to visit them.

Regardless of the actual way that a special promotion is published, it serves as a channel for local venues to advertise and attract more customers, which consequently can potentially translate to increased revenue.
There are anecdote stories for businesses that exploit such opportunities to their benefit.
For example, a burger joint in Philadelphia that offered a free beer with every Foursquare check-in is such a success story \cite{Fsq-success-stories}.
In fact our data verify that the specific venue (denoted as $\venue_P$) has benefited from Foursquare special offers.  
Nevertheless, conclusions drawn from similar bright examples are always affected by {\em sampling bias}.
Hence, the goal of our paper is to analyze and model the effectiveness of special offers through location-based social media/networks at scale.
This is the first work to analyze promotions offered through LBSNs.
We would like to emphasize here that our study is not focused on any specific platform (e.g., Foursquare).
On the contrary, our work is focused on the generic mechanism of promotions through LBSNs and our contribution is twofold:

{\bf (i)} We analyze the effectiveness of this mechanism using a large dataset we collected that includes time-series information from approximately 14 million venues on Foursquare.
Given that we do not have access to actual revenue data for venues our evaluation metric is the number of check-ins in a venue.  
We examine both periods during a promotion and after it is completed.
Our analysis combines randomization and statistical bootstrapping.
In brief, we use a randomly selected set of {\em matched} reference venues that have not offered any promotion during the data collection period to build a baseline for the probability of observing an increase in a venue's check-ins.  
We then compare this probability with the one computed using venues that have offered a promotion.
We further use block bootstrapping for non-parametric hypothesis testing to identify the venues for which the change observed in their check-ins is statistically significant.  Consequently, we obtain a more robust estimate of the aforementioned probabilities.  
%
{\bf Our main result indicates that the positive effects of special offers through LBSNs are more limited than what anecdote success stories might suggest. }
In particular, the probability of an increase in the mean daily check-ins for a venue that offers a promotion is approximately equal to that of the matched reference venues that do not offer any promotion.
Moreover, the standardized effect size on the daily check-ins is not very much different for the venues with promotions as compared to that of the reference venues.


{\bf (ii)}  We investigate whether there are specific factors that can drive success for a promotion.  
In particular, we build classifiers, by identifying relevant features, that can provide an educated decision on whether a specific venue will enjoy positive benefits through a special offer campaign.
The extracted features belong to three broad categories, that is, venue-related, promotion-related and geographical features.
Our experiments indicate that we can achieve good classification performance.
For instance, using simple models such as logistic regression we can achieve 83\% accuracy with 0.88 AUC.  
Interestingly, as we will elaborate on later, our model evaluations reinforce our findings from our statistical analysis, since the promotion-related features improve the classification performance only marginally.

\section{Our Dataset}
\label{sec:data}

We used Foursquare's public venue API during the period 10/22/2012-5/22/2013 and queried information 
for {\bf 14,011,045} venues once every day.
Each reading has the following tuple format: {\tt <ID, time, \# check-ins, \# users, \# specials, \# tips, \# likes, tip information, special information>}.
During the data collection period, there are 206,163 venues in total that have published at least one special offer.
Approximately 45\% of these venues publish only one special.
Furthermore, there are in total 735,034 unique special deals, with 88.68\% of them being provided by venues in the US.

\begin{figure}[h]
  \centering
  \includegraphics[scale=.45]{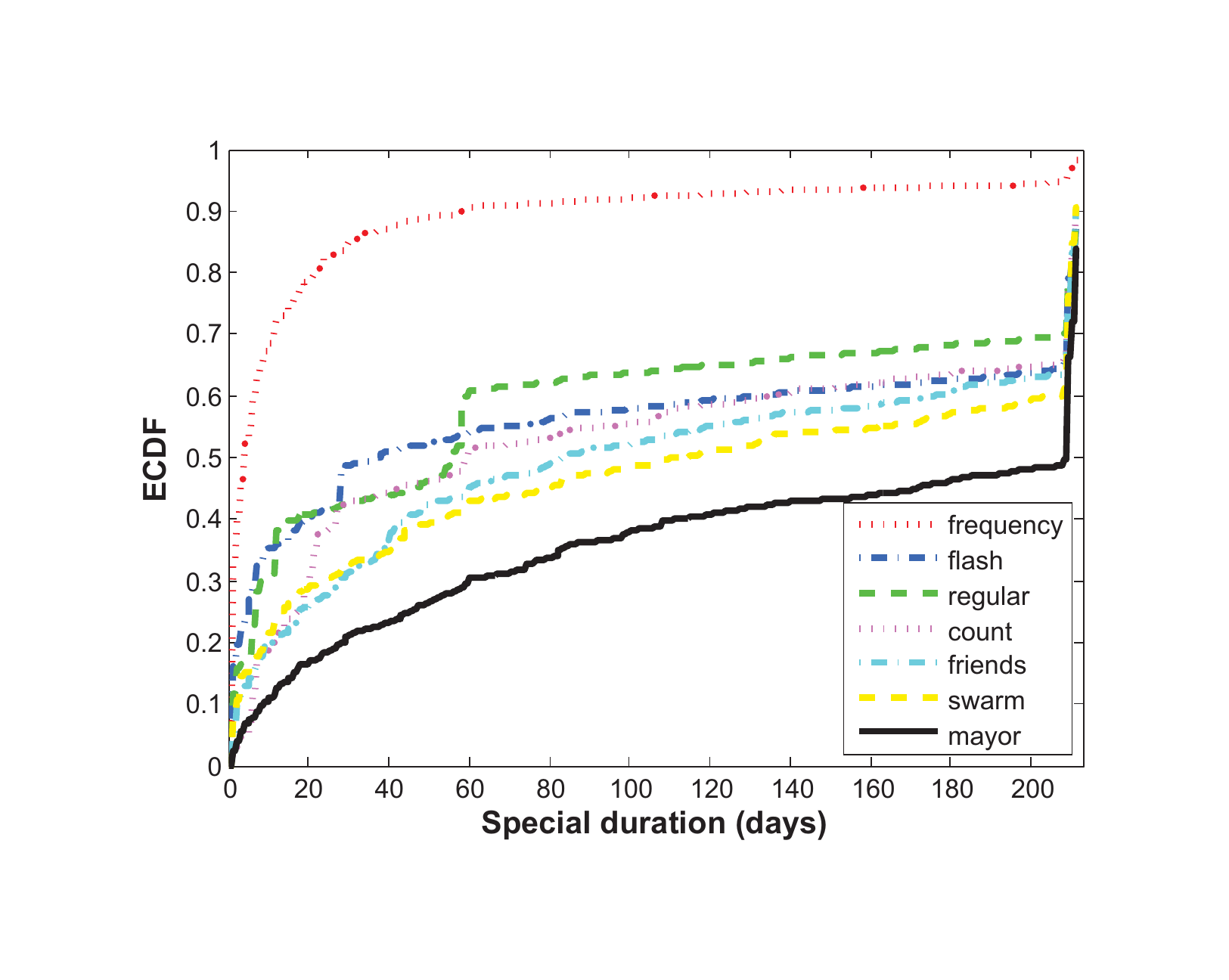} 
     \caption{``Frequency'' and ``Flash'' specials are usually shorter than other types of specials, while the ``Mayor'' special often lasts for a longer time.}
  \label{fg:cdf_special_duration}
\end{figure}

At the time, Foursquare had 7 types of specials, namely, ``Newbie'', ``Flash'', ``Frequency'', ``Friends'', ``Mayor'', ``Loyalty'' and ``Swarm'', each requiring different conditions to be earned \cite{fsq-specials}.
``Frequency'' is the most popular one in our dataset, possibly because compared to other types appears to be the easiest one to be {\em unlocked}, covering approximately 86.5\% of all the offers we collected.

Another parameter of interest for the special offers is their time duration.
Figure \ref{fg:cdf_special_duration} presents the empirical CDF of the offer duration. As we can see, ``Frequency'' and ``Flash'' special offers usually are active for a short duration, while ``Friends'' and ``Swarm'' usually last for a longer time possibly due to their stricter requirements. The ``Mayor'' special often lasts even longer, since a customer needs to become the Foursquare {\em mayor} of the venue to unlock the deal.
The {\em mayorship} is only awarded to the user who has the most check-ins in the venue during the last two months\footnote{The newest version of Foursquare does not include the notion of mayor anymore.}.

As alluded to above, a venue might offer multiple specials during the 7-month data collection period.
These multiple specials can be fully overlapped (i.e., they start and end at the same time), partially overlapped, or sequential.
We further define a {\bf promotion period} of a venue to be a continuous time period that the venue provides at least one offer and does not include more than two consecutive days without a special offer. 
In our dataset, approximately half of the promotions last for more than a week.
While a promotion as defined above can include multiple individual offers, for simplicity we will use the terms promotion, offer, campaign and deal interchangeably in the rest of the paper.

Finally, Foursquare associates each venue {\venue} with a category/type $\type(\venue)$ (e.g., cafe, school etc.). This classification is hierarchical.  
At the top level of the hierarchy there are 9 categories;
{\it Nightlife Spots}, {\it Food}, {\it Shops} \& {\it Services}, {\it Arts} \& {\it Entertainment}, {\it College} \& {\it University}, {\it Outdoors} \& {\it Recreation}, {\it Travel} \& {\it Transport}, {\it Residences} and {\it Professional} \& {\it Other Places}.
From these types, ``Food'', ``Nightlife Spots'' and ``Shops \& Services'' have the highest chances of offering a special deal (0.025, 0.04 and 0.016 respectively).
This can be attributed to the fact that the majority of the venues in these categories are commercial and hence, advertisement is most probably among their priorities.

\section{Effectiveness of Special Offers}
\label{sec:analysis}


{\bf Evaluation metric: }
Our data are in a time-series format and we also know the start ($t_s$) and the end ($t_e$) times of the promotion period.
Using these points we split each time-series to three parts that span the following periods: (i) {\bf before} the special campaign, $[t_0,t_{s-1}]$, (ii) {\bf during} the special campaign, $[t_s,t_e]$, and (iii) {\bf after} the special campaign, $[t_{e+1},t_n]$.
The key idea is to examine and analyze the changes that occur on the {\em daily check-ins} across these three time periods.

{\bf Data processing: }
Let us denote the original time-series collected for the check-ins in venue {\venue} with ${\checkinst}_{\venue}[t]$.
Simply put,  ${\checkinst}_{\venue}[t]$ is the accumulated number of check-ins in {\venue} at time $t$.
As aforementioned we obtain one reading every day for every venue.
However, consecutive readings might not be exactly equally-spaced in time due to a variety of reasons (e.g., network delays, API temporal inaccessibility etc.).
Hence, we transform each time series to the intended reference time-points using interpolation.
For the rest of the paper ${\checkinst}_{\venue}[\tau]$ will represent the interpolated time-series for the total number of check-ins in {\venue} with $\tau_{i+1}-\tau_i = 24$ hours.

We focus on campaign periods of venues in the US 
that last for at least 7 days and for which we have {\em enough} points in the time-series before the special offer (i.e., at least 4 weeks).
This allows us to build a representative baseline for the venue popularity prior to the promotion.
The above filters provide us with a final dataset of 40,071 promotion periods that we use in our analysis, offered by 36,567 venues.
We refer to this dataset as the {\em promotion} dataset.
Note here, that only a subset of those can be used for studying the long-term effect of the promotion.
In particular, for 26,355 of them we have enough points in the time-series after the special offer, and we use them for the long-term effect study.

Since our metric of interest is the daily check-ins, we utilize the first-order difference of the aggregated time series: 

\begin{equation}
{\norcheckinst}_{\venue}[\tau] = {\checkinst}_{\venue}[\tau] - {\checkinst}_{\venue}[\tau-1]
\label{eq:norcheckinst}
\end{equation}

The time-series we collected might exhibit biases that affect our analysis.
For instance, a change in a venue's daily check-ins might simply be a result of a change in the popularity of the social media application.
Moreover, seasonality effects can distort the contribution of the campaign on ${\norcheckinst}_{\venue}[\tau]$.
To factor in our analysis similar potential sources of bias we use a randomly selected, matched, reference group of venues that can account for the effects of similar externalities.

\subsection{Promotion dataset analysis}
\label{sec:promotions}

We begin by examining the fraction of promotions that enjoy an increase in the mean number of check-ins per day.
Let us denote the mean check-ins per day in venue {\venue} before the promotion (i.e., during the period $[t_{s-k},t_{s-1}]$) with $\median_{\norcheckinst_\venue}^b$.
We similarly define the average check-ins per day in {\venue} during (i.e., in the time period $[t_{s},t_e]$)
and after (i.e., in the time period $[t_{s+1}, t_{s+w}]$)
the promotion campaign as $\median_{\norcheckinst_\venue}^d$ and $\median_{\norcheckinst_\venue}^a$ respectively.
To reiterate, in order to build a concrete baseline for the period prior to the promotion we set $k = 28~days$.
In order to study the long term effect of the promotion we would like to have a stabilized time interval after the campaign is over.
Hence, we include in our analysis only the venues for which we have data for at least 7 days after the end of the promotion.
Consequently, we set $w=k$, if we have 28 days of data after the promotion.
Otherwise we set $w$ equal to the number of time-points available (i.e., $7 \le w \le 28)$.

\begin{figure*}[t]
\centering
\subfloat[Short-term]{
   \includegraphics[scale =0.3] {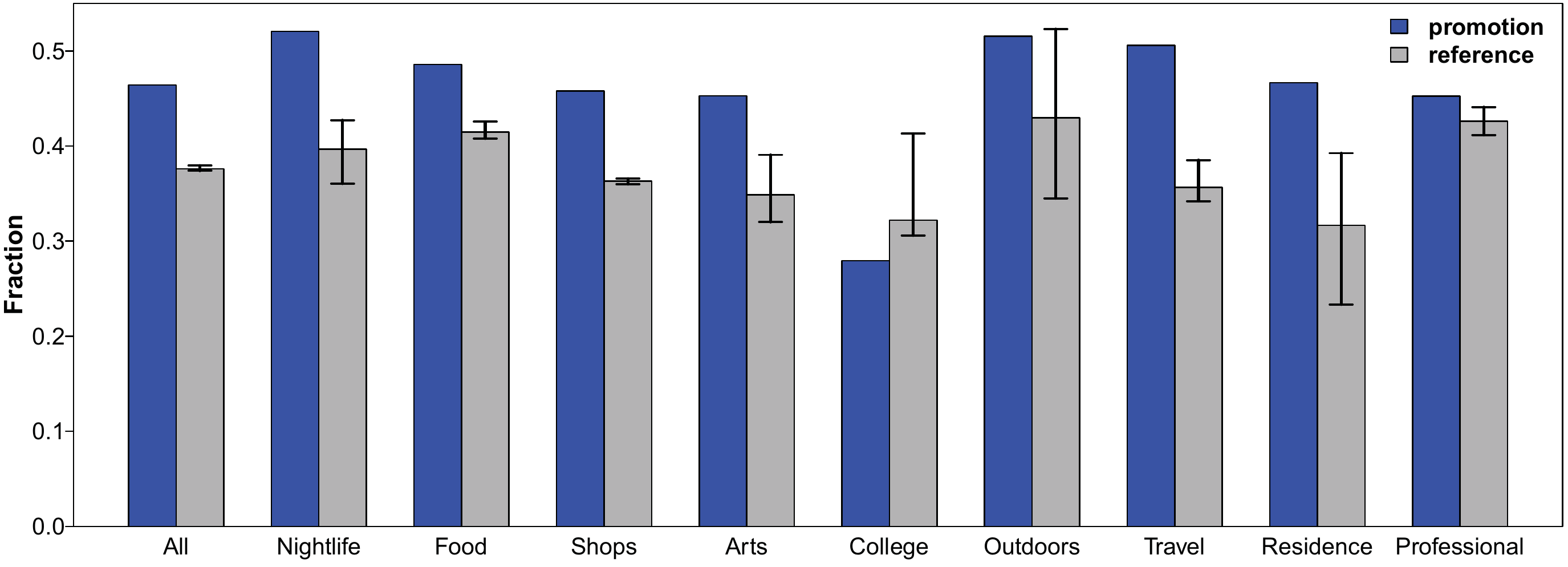}
   \label{fig:positive_percentage_during}
 }
 \subfloat[Long-term]{
   \includegraphics[scale =0.3] {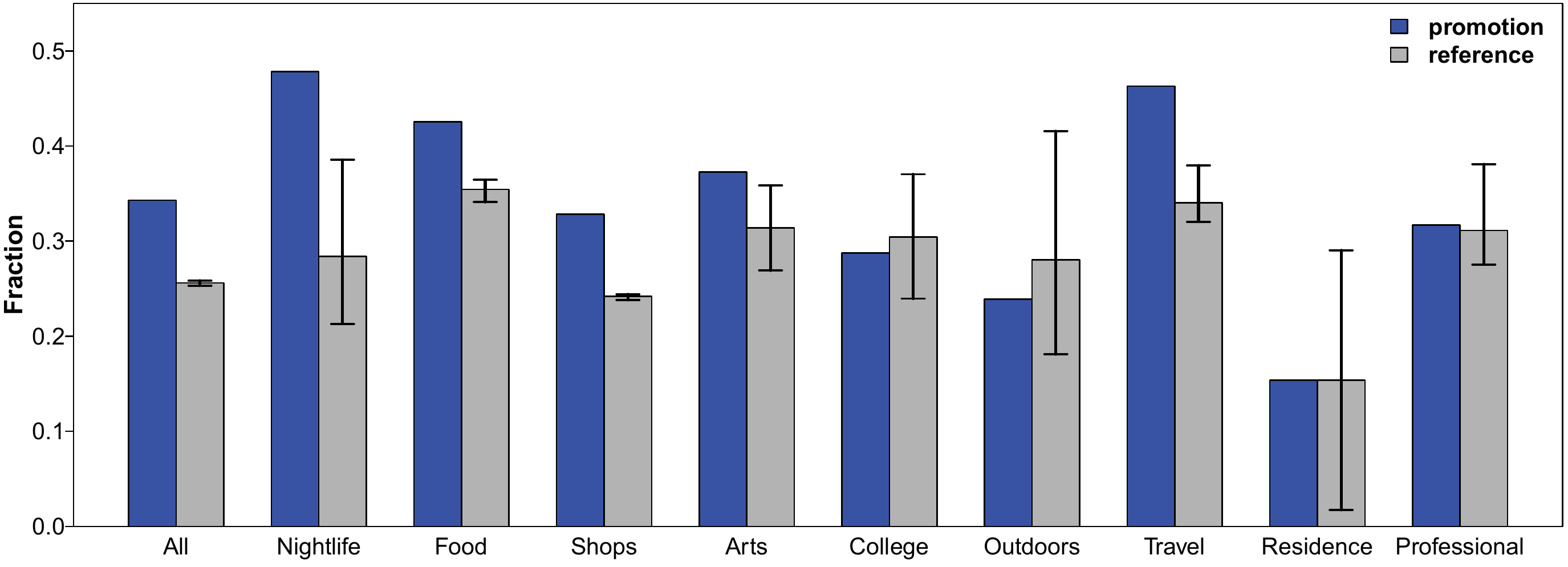}
   \label{fig:positive_percentage_after}
 }
\caption{Fraction of venues exhibiting an increase in the mean daily check-ins.}
\label{fig:positive_percentage}
\end{figure*}

Given this setting we first compute the difference $\median_{\norcheckinst_\venue}^d - \median_{\norcheckinst_\venue}^b$ ($\median_{\norcheckinst_\venue}^a - \median_{\norcheckinst_\venue}^b$).
A positive sign essentially translates to an increase in the average daily check-ins during (after) the promotion period.
Figure \ref{fig:positive_percentage} depicts our results.
As we can see, the fraction of venues in the promotion group that enjoy an increase in their check-ins during the promotion is approximately 50\%, while a smaller fraction (about 35\%) exhibits an increase after the offer is ceased.
There is also some variation observed based on the venue type, with some categories exhibiting a larger fraction of venues with an increase (e.g., nightlife).
However, part of this variability might be attributed to the fact that for some categories we have a very small sample in the promotion set (e.g., we only have 128 promotions in {\it Outdoors} and 30 in {\it Residence}).  

\begin{figure*}[t]
\centering
\subfloat[All categories]{
   \includegraphics[scale =0.45] {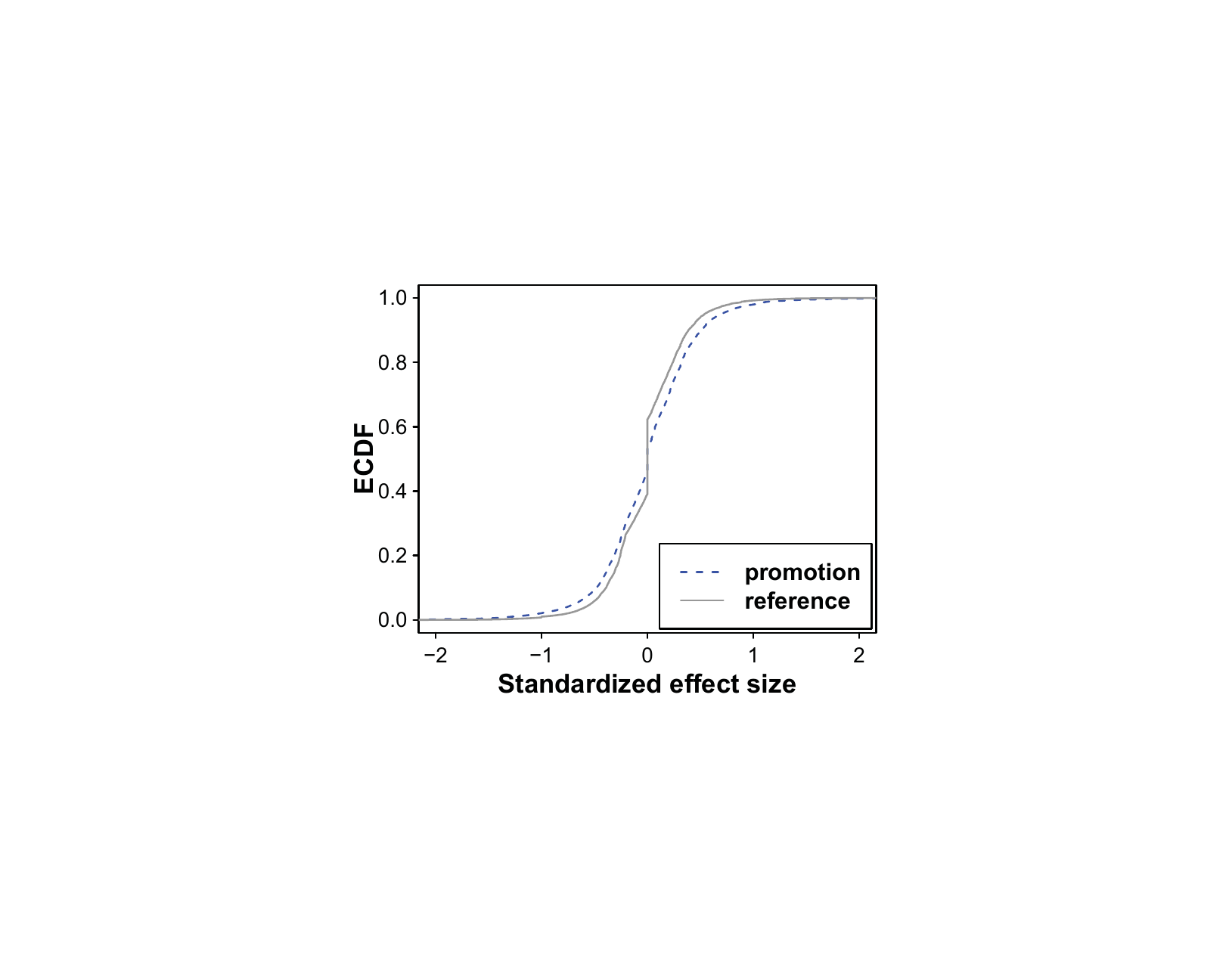}
   \label{fig:ECDF_effSize_ccnt_during}
 }
 \subfloat[Nightlife]{
   \includegraphics[scale =0.45] {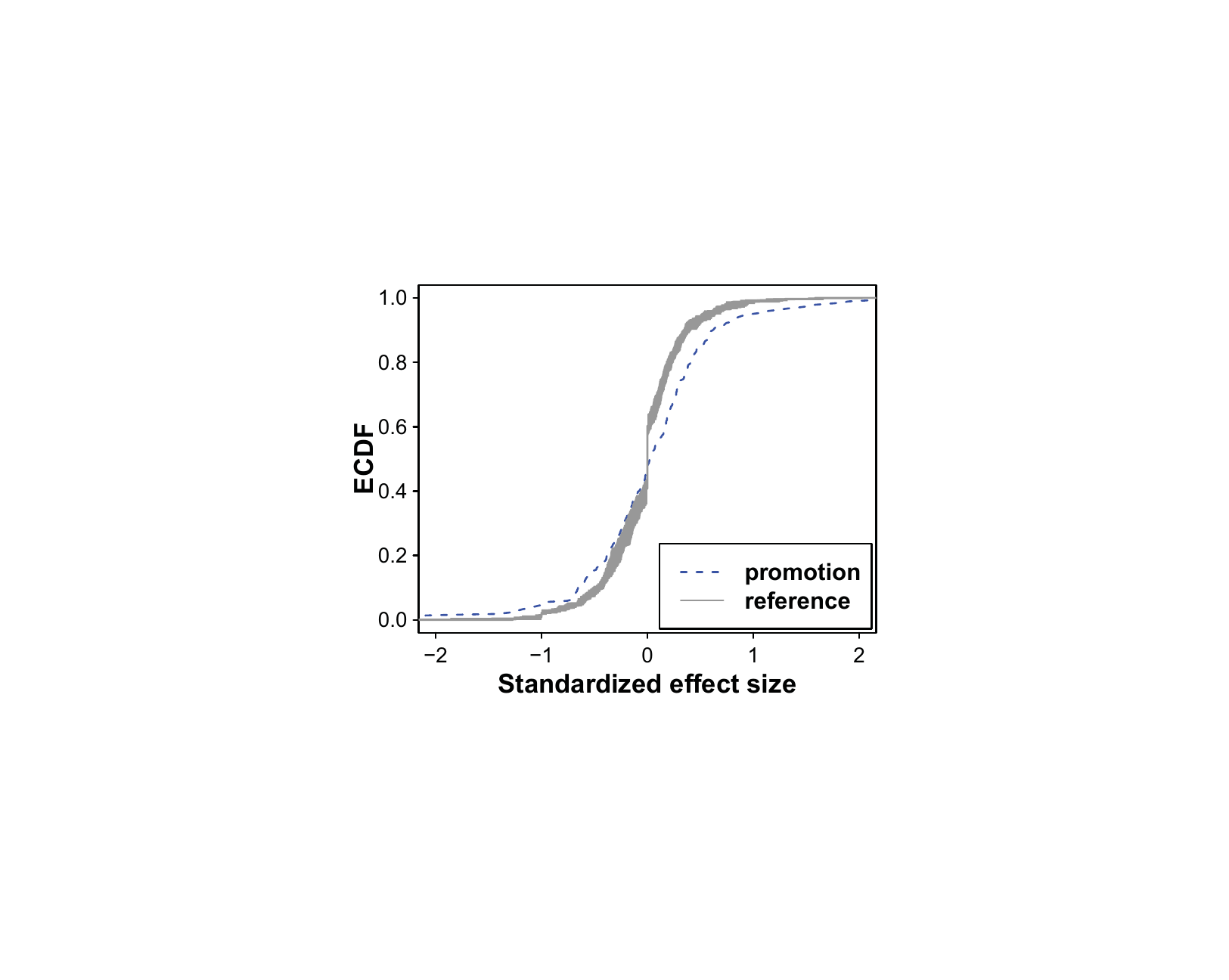}
   \label{fig:ECDF_effSize_ccnt_during_Nightlife}
 }
 \subfloat[Food]{
   \includegraphics[scale =0.45] {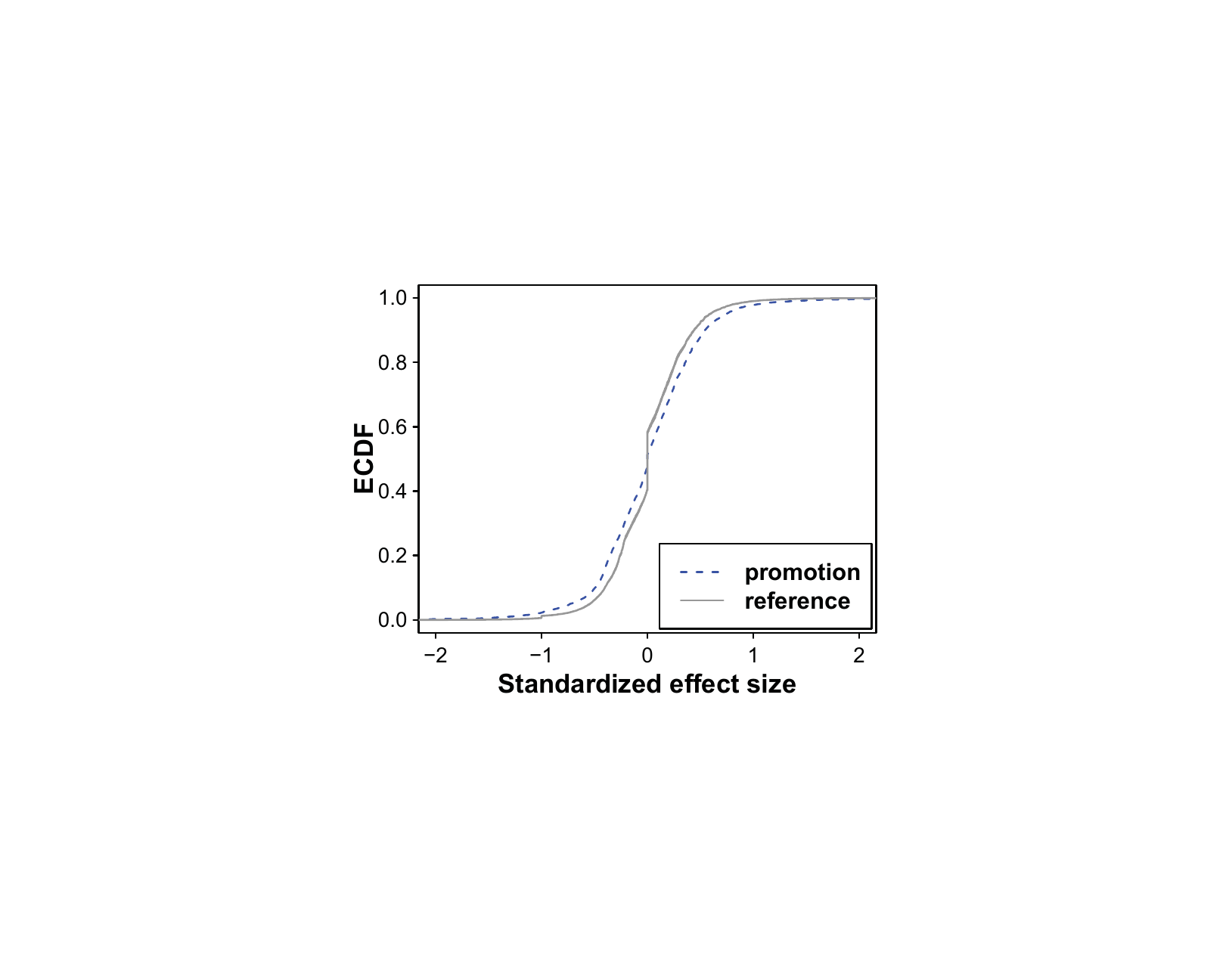}
   \label{fig:ECDF_effSize_ccnt_during_Food}
 }
 \subfloat[Shops]{
   \includegraphics[scale =0.45] {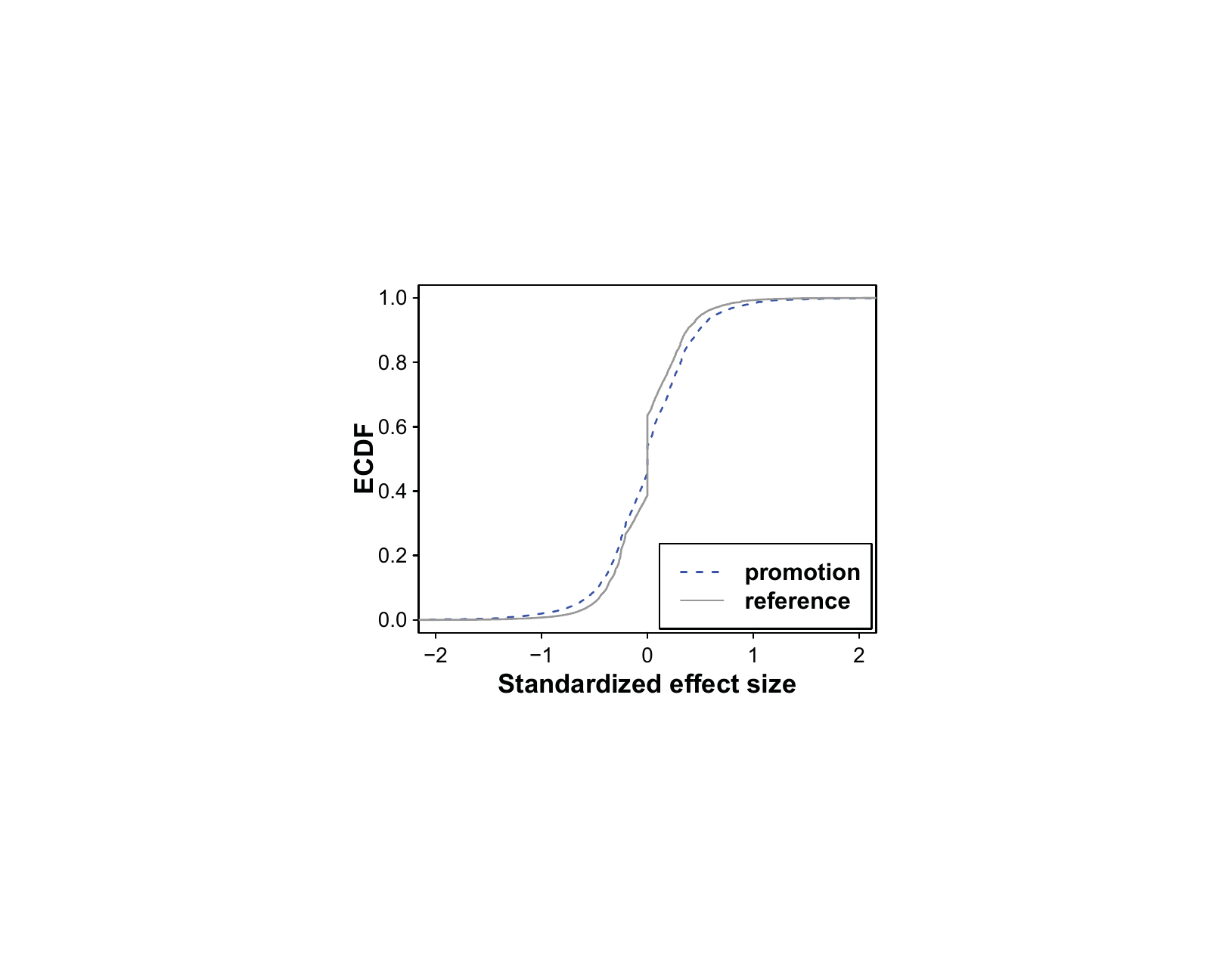}
   \label{fig:ECDF_effSize_ccnt_during_Shops}
 }
 \subfloat[Arts]{
   \includegraphics[scale =0.45] {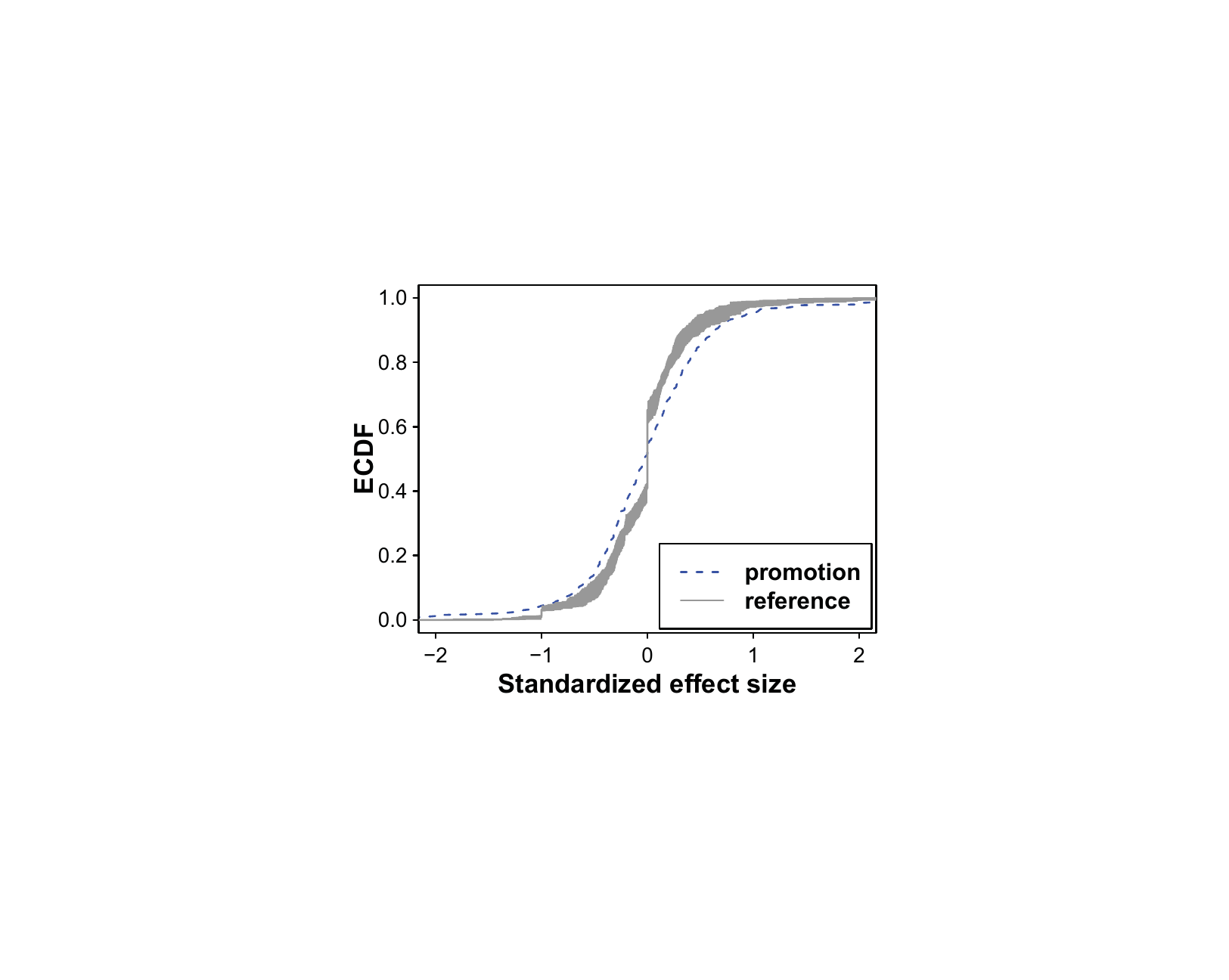}
   \label{fig:ECDF_effSize_ccnt_during_Arts}
 }
 \\
 \subfloat[College]{
   \includegraphics[scale =0.45] {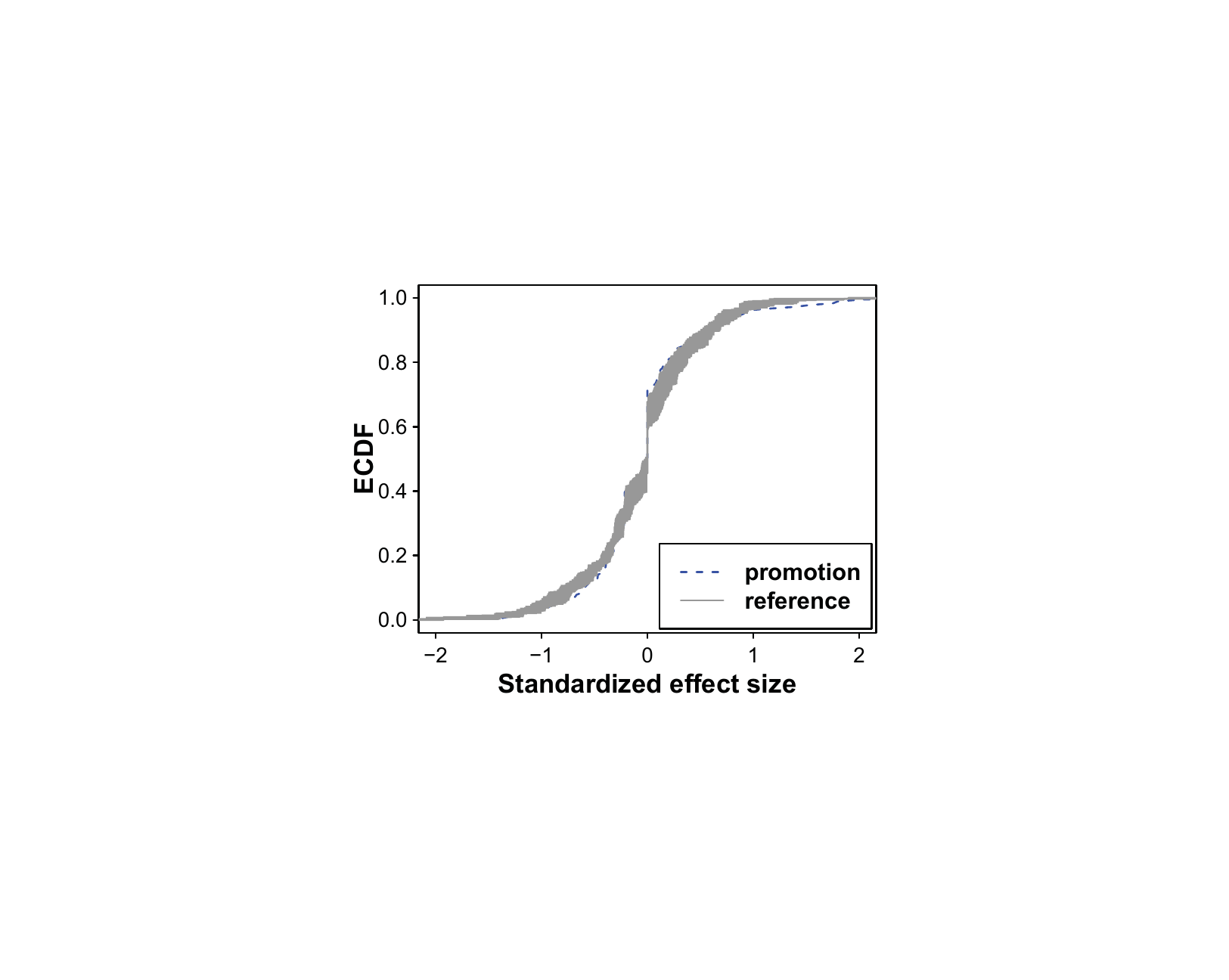}
   \label{fig:ECDF_effSize_ccnt_during_College}
 }
 \subfloat[Outdoors]{
   \includegraphics[scale =0.45] {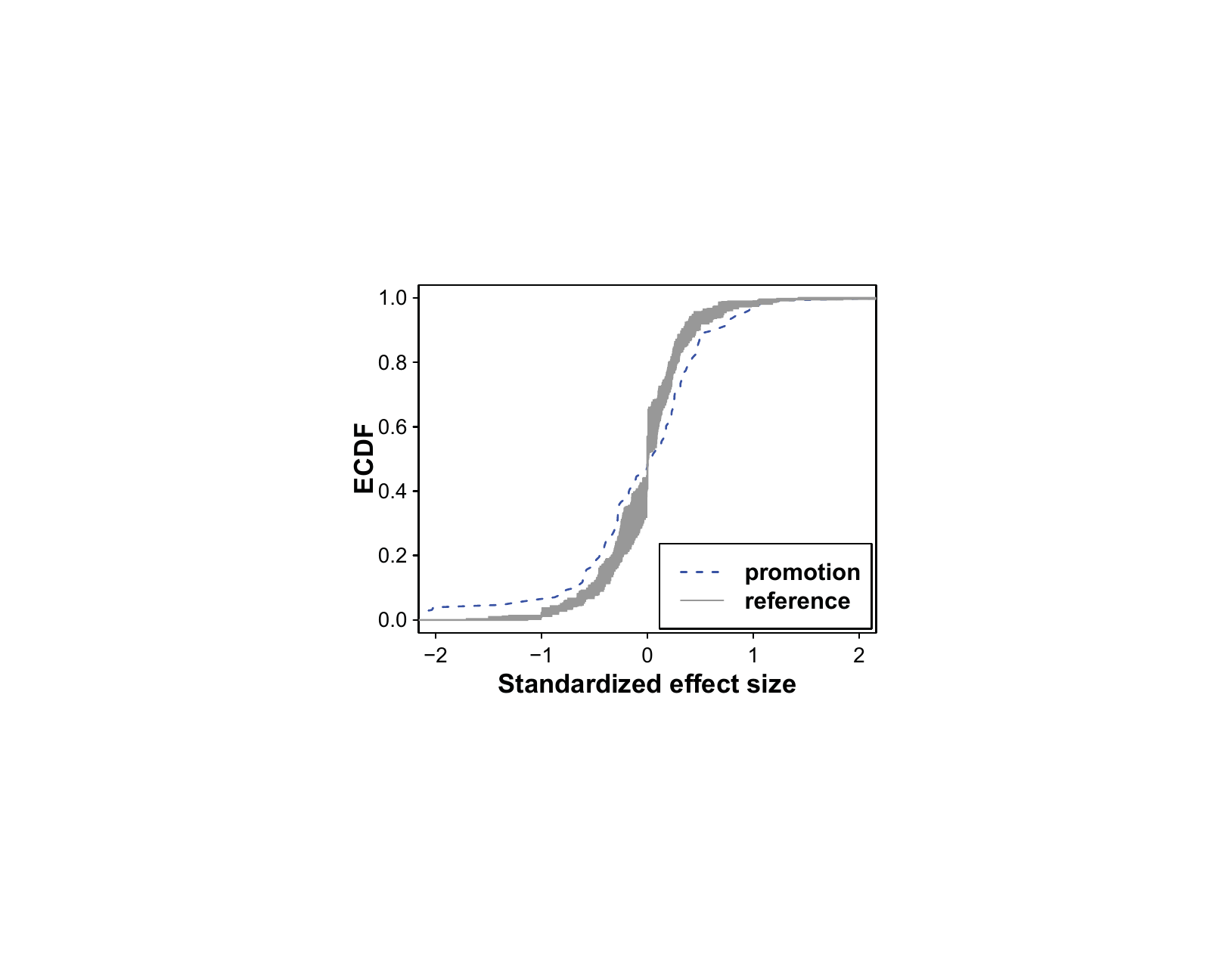}
   \label{fig:ECDF_effSize_ccnt_during_Outdoors}
 }
 \subfloat[Travel]{
   \includegraphics[scale =0.45] {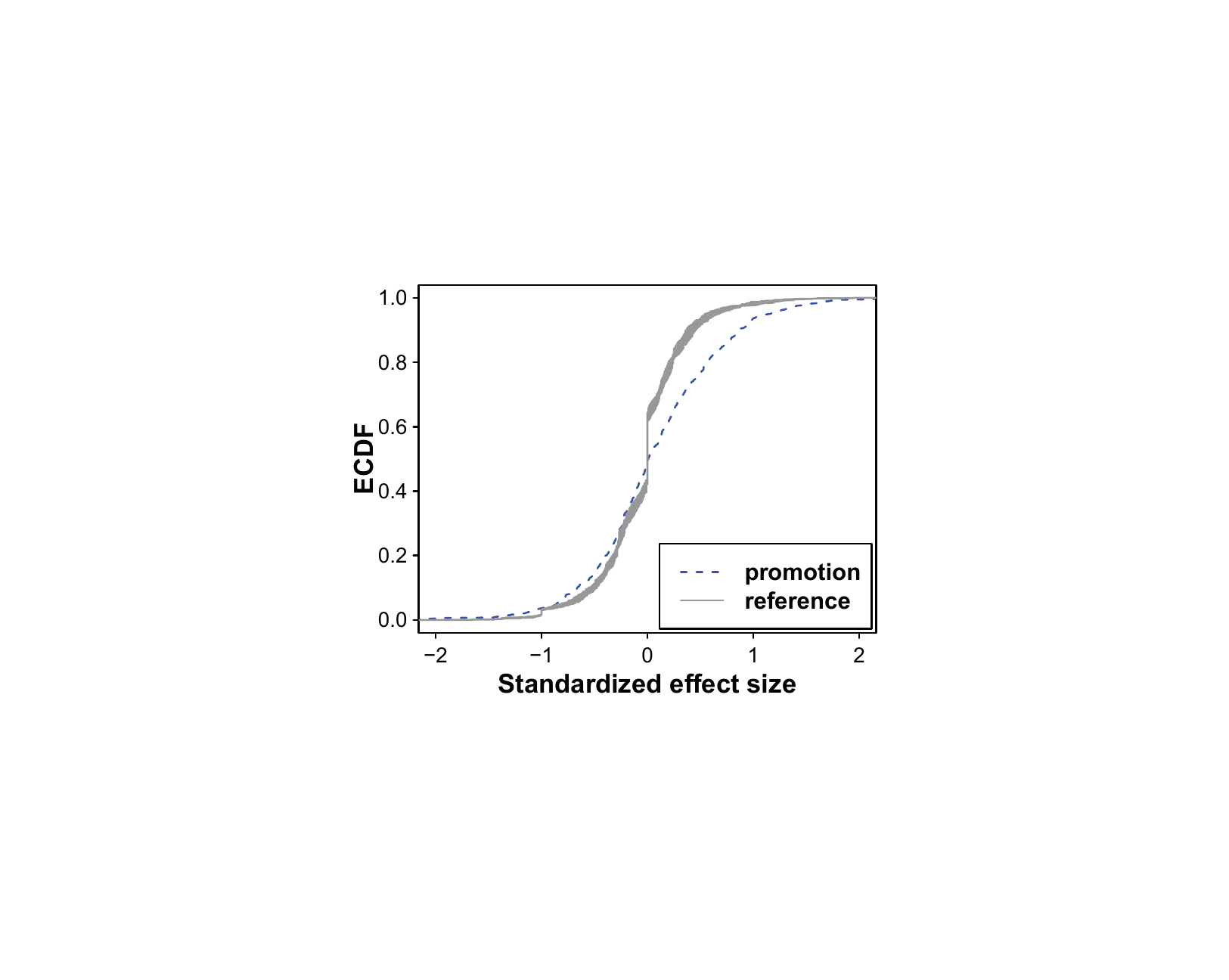}
   \label{fig:ECDF_effSize_ccnt_during_Travel}
 }
 \subfloat[Residence]{
   \includegraphics[scale =0.45] {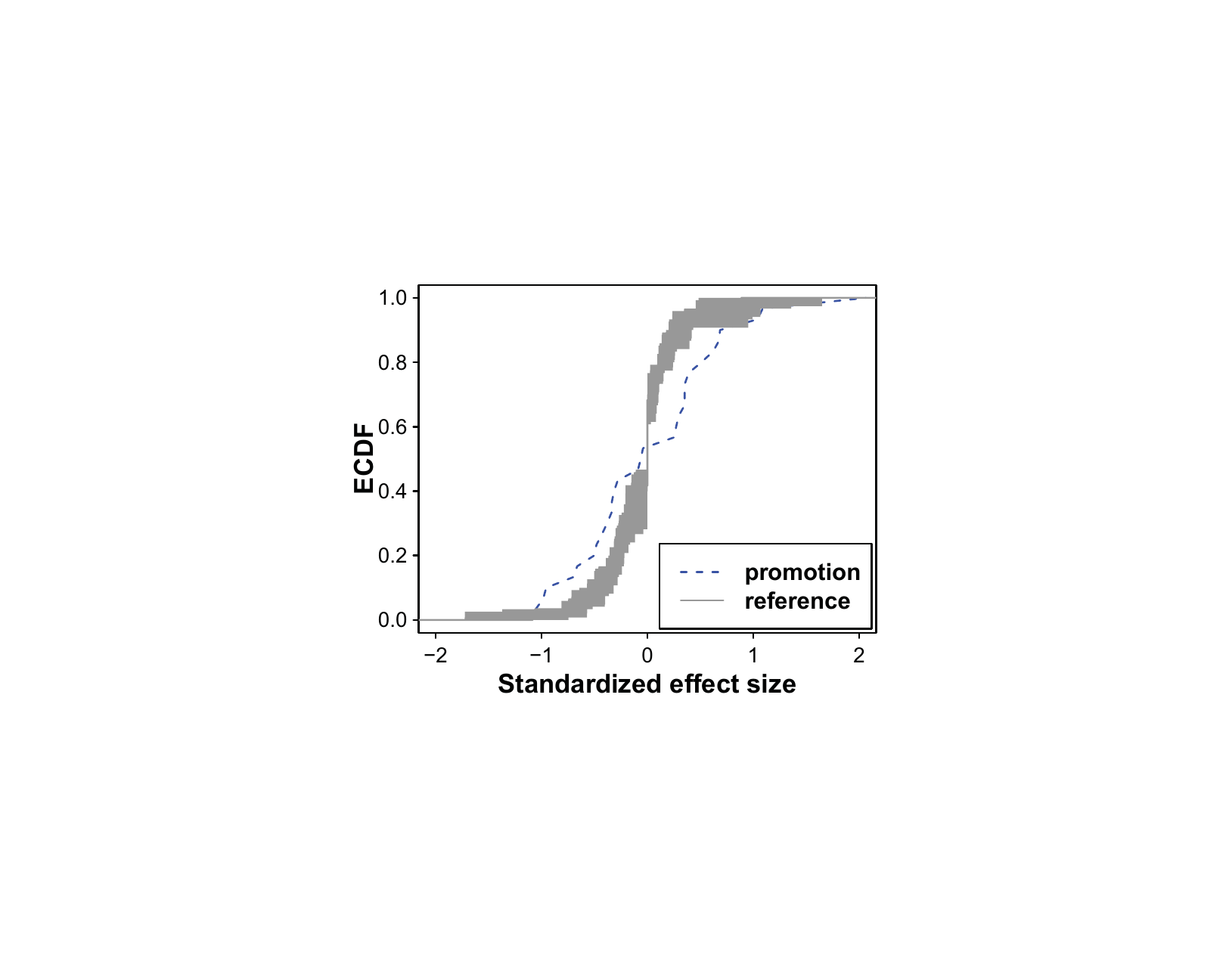}
   \label{fig:ECDF_effSize_ccnt_during_Residence}
 }
 \subfloat[Professional]{
   \includegraphics[scale =0.45] {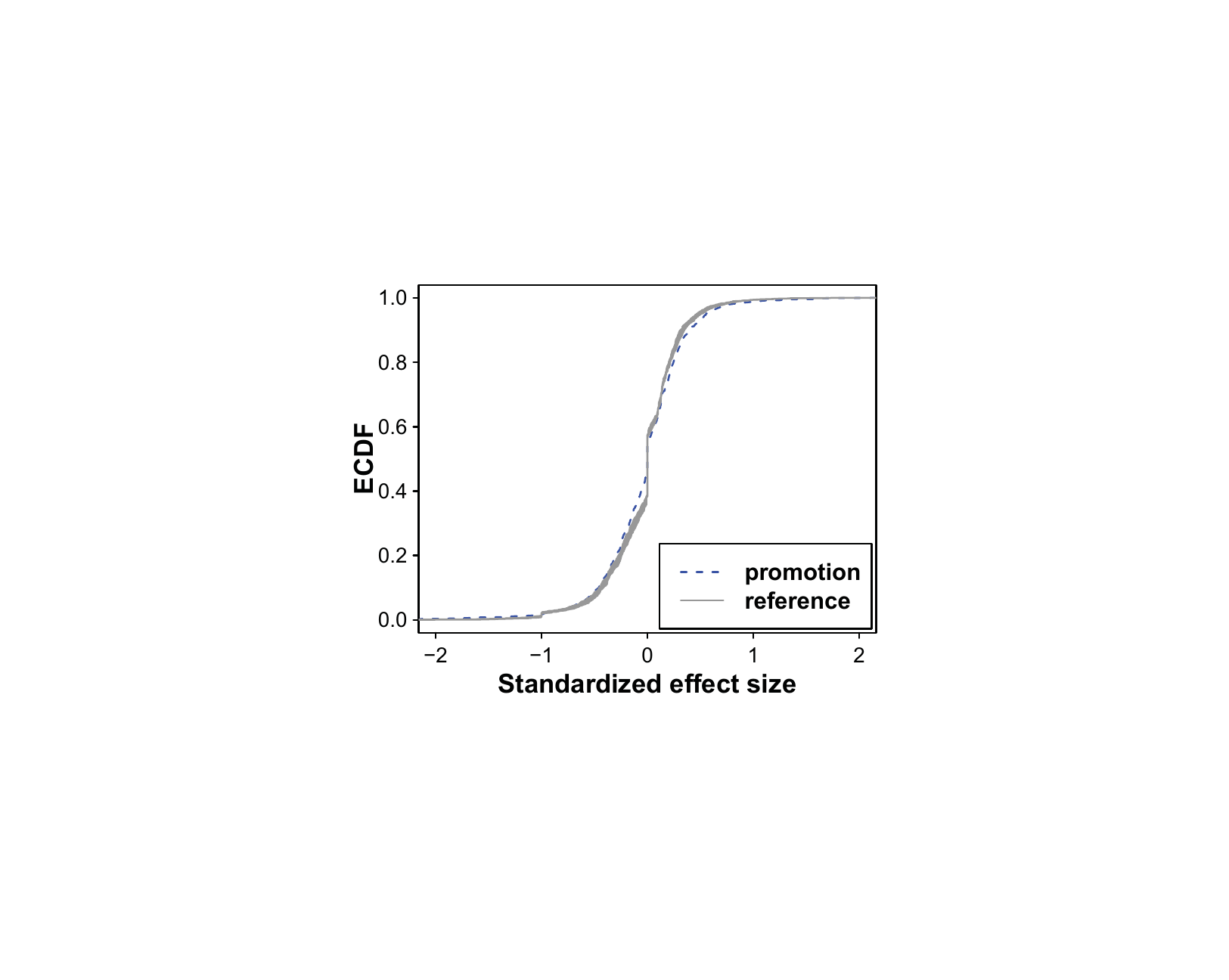}
   \label{fig:ECDF_effSize_ccnt_during_Work}
 }
\caption{Both the promotion and reference groups enjoy similar effect sizes.}
\label{fig:ecdf_during}
\end{figure*}

In summary, a large fraction of venues exhibit increase in their check-ins during and after the special offer.
However, an almost equal proportion of venues does not enjoy an increase in the average daily check-ins.
Next we delve further into the details of the effectiveness of local promotions.

\subsection{Reference venues}
\label{sec:randomized}

Our results above clearly cannot establish any causal relation between promotion campaigns and observed changes in the daily check-ins.
This would require careful design of field experiments.
However, it is not possible in our work since we only have access to observational data.
The direct comparison between venues that offer promotions and those that do not, can be affected by a {\em self-selection} bias of the promotion venues; venue owners might not randomly decide whether to offer a deal, but other confounding factors might affect this decision.

Therefore, in order to account for these confounding factors and other externalities, we opt to get a baseline for comparison by utilizing techniques for quasi-experimental studies. 
In particular, we randomly sample a reference group from the set of venues with no promotion, such that the distribution of specific {\em observed} features of this sample {\em matches} that of the promotion group.
This of course assumes that there is no selection bias based on unobserved characteristics.
The features 
we use for matching are the location as well as the type of the venue.
The reference group also ensures that on average the venues at both groups will experience similar externalities (e.g., seasonal effects, effects related to the popularity of Foursquare etc.).
Once the reference group is obtained, we sample the empirical promotion period distribution of the promotion venues and assign pseudo-promotion periods to the reference group venues.
Consequently we perform the same analysis described in the previous section on the reference group.

Our results from 20 non-overlapping reference groups are also depicted in Figure \ref{fig:positive_percentage}, where the 95\% confidence intervals are also presented.
As we can see the fraction of venues enjoying an increase in the promotion group is higher compared to that in the reference group.
If we denote with ${\increaseevent}_d$ (${\increaseevent}_a$) the event of an increase for $\median_{\norcheckinst_\venue}^d$ ($\median_{\norcheckinst_\venue}^a$), with
$\specialevent$ the event of a venue offering a special deal and with $\externalitiesevent$ the various environmental externalities that are present, the reference group opts to obtain an estimate for the probability $P({\increaseevent}_d |  \externalitiesevent)$.
On the other hand, the promotion group includes an additional externality, the presence of a promotion.
Hence, with the promotion group we are able to estimate $P({\increaseevent}_d |  \specialevent, \externalitiesevent)$.
Our results indicate that $P({\increaseevent}_d |  \specialevent, \externalitiesevent) > P({\increaseevent}_d |  \externalitiesevent)$ and $P({\increaseevent}_a |  \specialevent, \externalitiesevent) > P({\increaseevent}_a |  \externalitiesevent)$ when considering all types of venues.
However, the difference between these two probabilities is only about 0.1 for both the short and long term.

Another aspect related with the potential effectiveness of the promotion campaign is the actual effect size of the observed change.
The degree of this change can be captured through the standardized effect size of Cohen's $d$:

\begin{equation}
d = \frac{\median_{\norcheckinst_\venue}^d-\median_{\norcheckinst_\venue}^b}{\sigma_{pooled}}
\label{eq:cohen}
\end{equation}
where $\sigma_{pooled}$ is the pooled standard deviation of the two samples (before and during the promotion).
Figure \ref{fig:ecdf_during} 
presents the empirical CDF for the observed standardized effect sizes in both the promotion and the reference groups for the short term.  
The results for the long term are similar and omitted due to space limitations.
For the reference groups we also present the 95\% confidence intervals of the distributions.
As we can observe there is a shift in the distribution for the promotion group, which is different for different categories. 
However, this shift is very small.
Furthermore, an interesting point to observe is the jump at the reference groups' ECDF at $d=0$.
This means that there is a non-negligible fraction of venues in the reference group that have exactly the same mean for the two periods compared.
We come back to this observation in the following section.  

\subsection{Bootstrap tests}
\label{sec:bootstrap}

Our results above indicate that a large number of venues exhibit small effect sizes, which might not represent robust observations.
Therefore, in this section we opt to identify and analyze the promotions in our dataset that are associated with a statistically significant change in their check-ins.

Given our setting, the following two-sided hypothesis test examines whether their is a statistically significant change observed in the short-term:

\begin{eqnarray}
H_0: & \median_{\norcheckinst_\venue}^b = \median_{\norcheckinst_\venue}^d \label{eq:median_test_null}\\
H_1: & \median_{\norcheckinst_\venue}^b \neq \median_{\norcheckinst_\venue}^d \label{eq:median_test_alt}
\end{eqnarray}

If the $p$-value of the test is less than $\alpha$, then there is strong statistical evidence that we can reject the null-hypothesis (at the significance level of $\alpha$).
The sign of the observed difference will further inform us if the change is positive.  
In our analysis we pick the typical value of $\alpha = 0.05$.
If we want to examine the long-term effectiveness of special deals we devise the same test as in Equations (\ref{eq:median_test_null}) and (\ref{eq:median_test_alt}), where we substitute $ \median_{\norcheckinst_\venue}^d$ with $ \median_{\norcheckinst_\venue}^a$.
We choose to rely on bootstrap for the hypothesis testing rather than on the t-test to avoid any assumption for the distribution of the check-ins.
Bootstrap also allows us to estimate the statistical power $\pi$ of the performed test.
This is important since an underpowered test might be unable to detect statistically significant changes especially if the effect size and/or the sample size are small.
Consequently, this can lead to underestimation of the cases where the alternative hypothesis is true.

Statistical bootstrap \cite{efron93} is a robust method for estimating the unknown distribution of a population's statistic  when a sample of the population is known.
The basic idea of the bootstrapping method is that in the absence of any other information about the population, the observed sample contains all the available information for the underlying distribution.
Thus, resampling with replacement is the best guide to what can be expected from the population distribution had the latter been available.
Generating a large number of such resamples allows us to get a very accurate estimate of the required distribution.
Furthermore, for time-series data, block resampling retains any dependencies between consecutive data points \cite{kunsch89}.

In our study we will use block bootstrapping with a block size of 2 to perform the hypothesis tests.  
When performing a statistical test we are interested in examining whether under the null hypothesis, the observed value for the statistic of interest was highly unlikely to have been observed by chance.
In our setting, under $H_0$ the two populations have the same mean, i.e., $\median_{\norcheckinst_\venue}^d-\median_{\norcheckinst_\venue}^b=0$.
Hence, we first center both samples, before and during the special, to a common mean (e.g., zero by subtracting each mean respectively) in order to make the null hypothesis true.
Then we bootstrap each of these samples and calculate the difference between the new bootstrapped samples.
By performing {\boots} = 4999 bootstraps, we are able to build the distribution of the difference $\median_{\norcheckinst_\venue}^d-\median_{\norcheckinst_\venue}^b$ under $H_0$.
If the ($1-\alpha$) confidence interval of $\median_{\norcheckinst_\venue}^d-\median_{\norcheckinst_\venue}^b$ under the null hypothesis does not include the observed value from the data, then we can reject $H_0$.
An empirical $p$-value can also be calculated by computing the fraction of bootstrap samples that led to an absolute difference greater than the one observed in the data.

With statistical bootstrapping we can further estimate the power $\pi$ of the statistical test performed.  $\pi$ is the conditional probability of rejecting the null hypothesis given that the alternative hypothesis is true.
For calculating $\pi$ we start by following exactly the same process as above, but without centering the samples to a common mean.
This will allow us to build the distribution of $\median_{\norcheckinst_\venue}^d-\median_{\norcheckinst_\venue}^b$ under $H_1$.
Then the power of the test is the overlap between the critical region and the area below the distribution curve under $H_1$.

\begin{figure}
  \centering
  \includegraphics[scale=0.45]{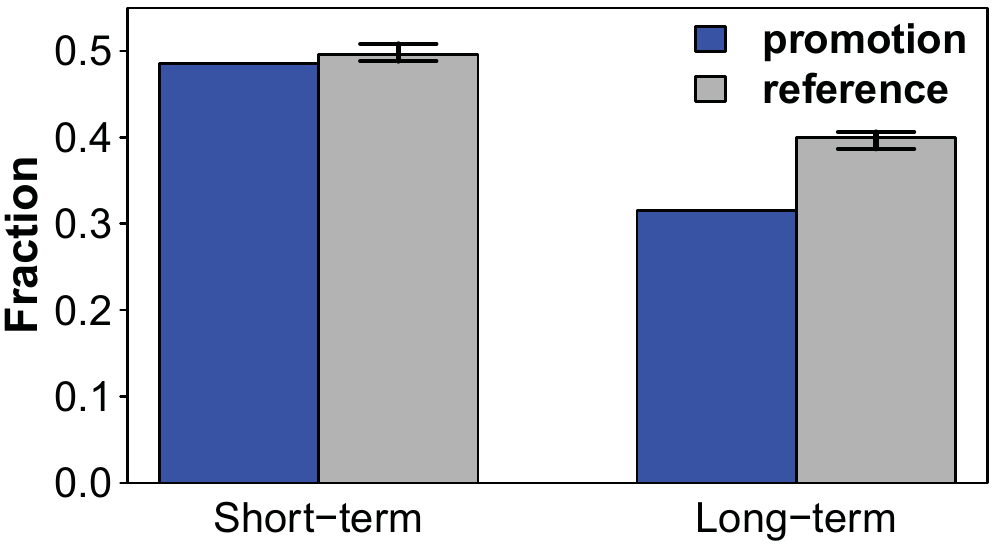}\\
  \caption{When considering venues with robust changes in their check-ins the effect of local promotions disappear.}
  \label{fig:bootstrap_binary_distribution}
\end{figure}

We have applied the bootstrap hypothesis test on our promotion and reference groups.
Figure \ref{fig:bootstrap_binary_distribution} presents our results for all types of venues. Similar behavior is observed for specific venue categories. However, due to space limitations, the results are omitted.
In particular, we calculate the fraction of promotions associated with a statistically significant increase in the average daily check-ins.
Note that we consider only the promotions whose $p$-value is less than $\alpha= 0.05$ or $\pi \ge 0.8$ (the latter is a typical value used and increases our confidence that failure to reject $H_0$ was not due to an underpowered test).
As we can see, in this case the fraction of venues that exhibit an increase in the average daily check-ins is the same for both groups, i.e., $P({\increaseevent}_d |  \specialevent, \externalitiesevent) \approx P({\increaseevent}_d |  \externalitiesevent)$.
This suggests that the presence of a local promotion and the increase in the average check-ins are conditionally independent given the externalities $\externalitiesevent$!
For the long term we see a smaller fraction of promotion venues enjoying a positive change in their check-ins.
While the reasons for this are not clear, recent literature has reported similar findings in a tangential context.
Byers {\em et al.} \cite{byers2012daily} found that venues offering Groupon deals see a reduction in their Yelp ratings after the promotion.
Along the same lines, Foursquare venues that offer promotions appear more probable to see a reduction in their daily check-ins.
Unfortunately, more than half of the venues in our datasets are not rated and hence, we cannot directly examine the effect of promotions on the rating.

\begin{figure}[t]
\centering
\subfloat[Short-term]{
   \includegraphics[scale =0.4] {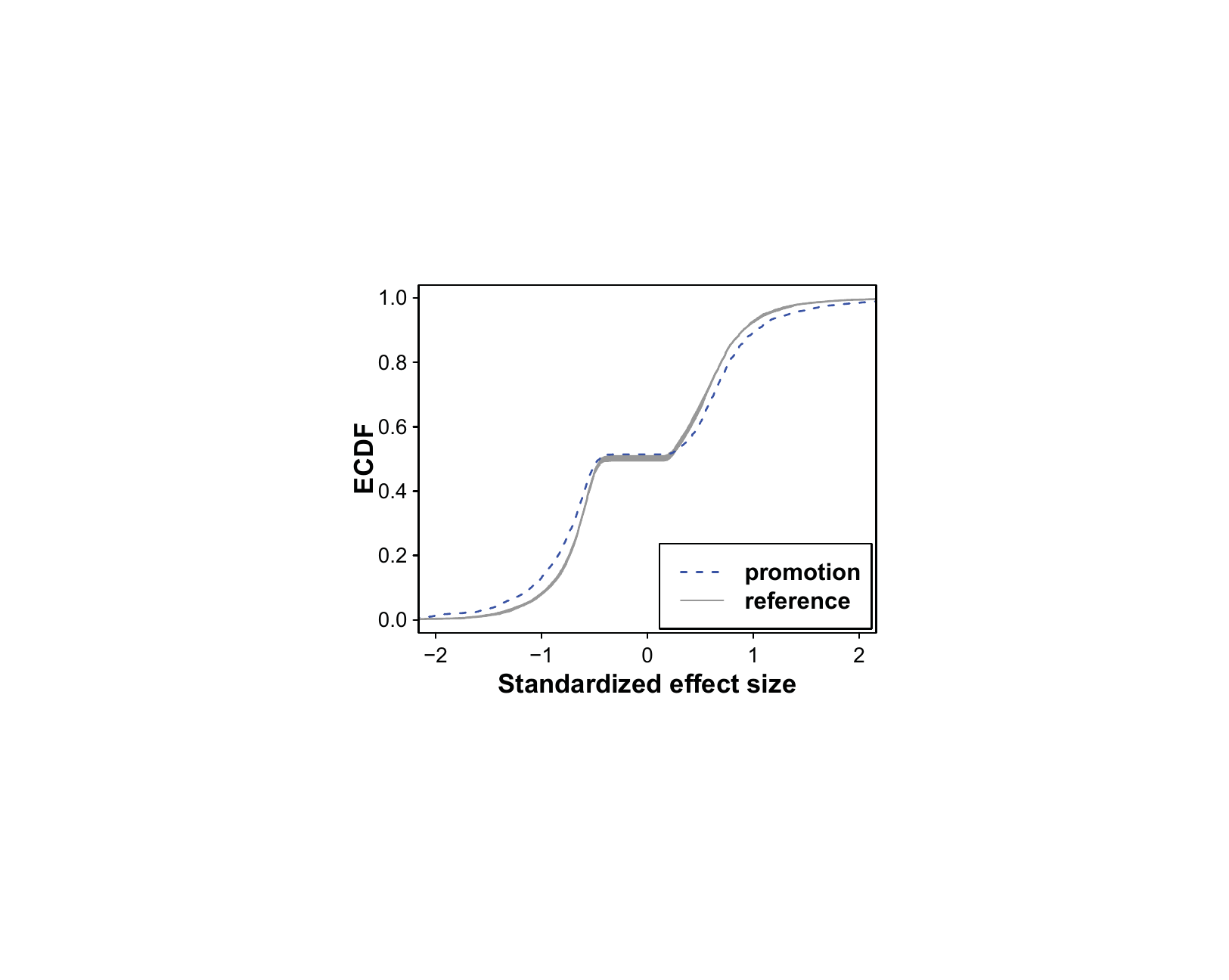}
   \label{fig:bootstrap_ECDF_effSize_ccnt_during}
}
~~~~~
\subfloat[Long-term]{
   \includegraphics[scale =0.4] {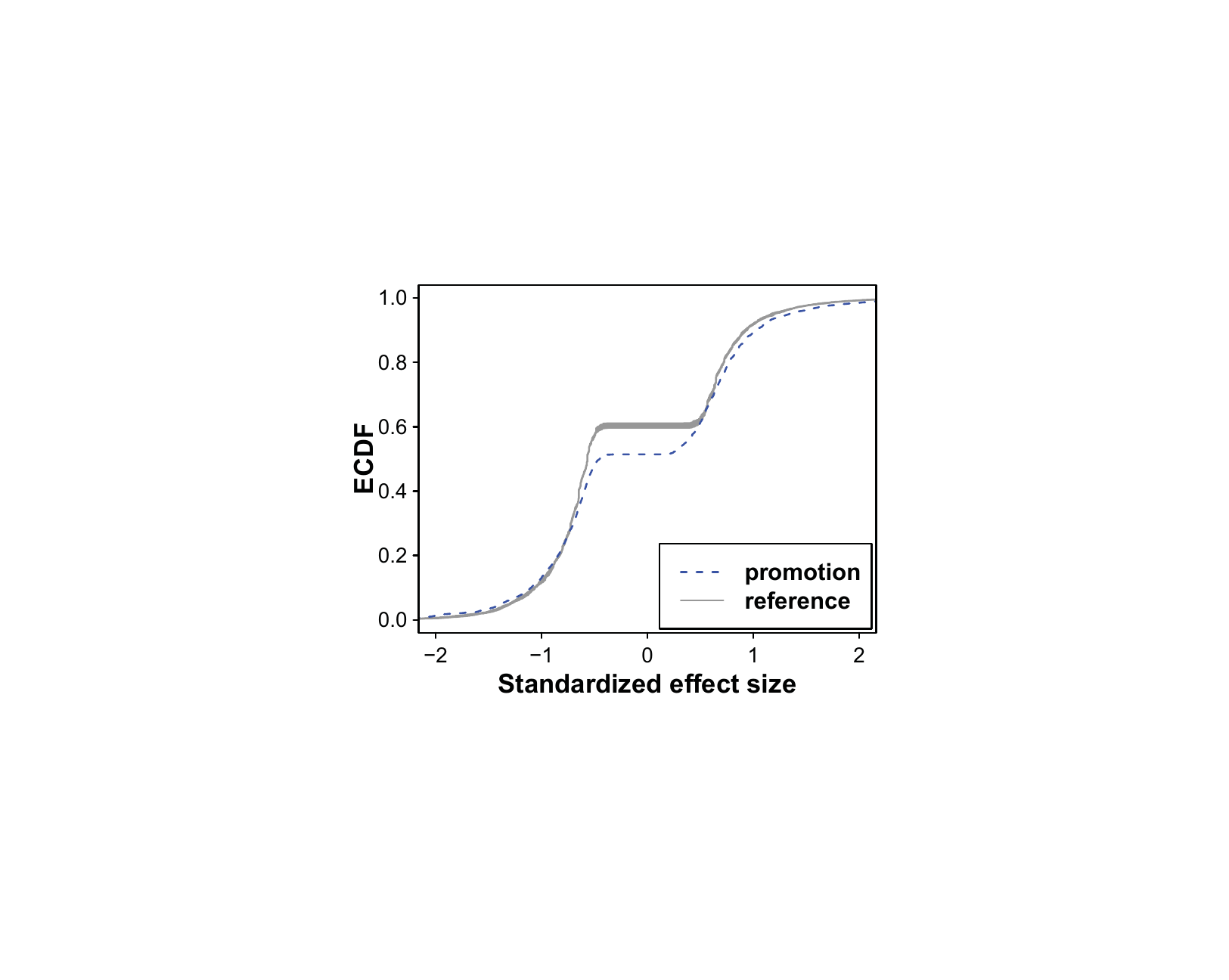}
   \label{fig:bootstrap_ECDF_effSize_ccnt_after}
 }
\caption{Small effect sizes do not provide robust observations based on our bootstrap tests.}
\label{fig:bootstrap_ECDF_effSize}
\end{figure}

More importantly though, in the previous section we emphasized on the fact that the reference group includes a large proportion of venues with $d=0$.
This clearly reduces the fraction of venues in the reference groups that have $d>0$ leading to smaller bars for the reference group in Figure \ref{fig:positive_percentage}.
A further examination of these cases shows that the vast majority of these venues exhibit 0 check-ins over the whole period.
These data points do not represent real venues, but are venues that correspond to events such as extreme weather phenomena, traffic congestion, potentially spam venues etc.
Hence, we can remove these venues from our reference groups.
After doing so we are able to recover the results presented in Figure \ref{fig:bootstrap_binary_distribution} further supporting the conditional independence between an increase in the mean number of check-ins per day and promotions.
Note that our bootstrap tests for these venues are extremely underpowered (practically there is not any distribution since every observation is 0) and hence, are not included in the results presented in Figure \ref{fig:bootstrap_binary_distribution}.
As we can further see from the plateau around $d=0$ in Figure \ref{fig:bootstrap_ECDF_effSize} that depicts the empirical CDF of Cohen's $d$ for the venues used in Figure \ref{fig:bootstrap_binary_distribution}, small effect sizes do not constitute {\em robust} observations.
Of course this can either be due to the low power of the test to detect a small effect size, or due to the actual non-existence of any effect.

\subsection{Anecdote evaluation}
\label{sec:anecdote}

\begin{figure}[t]
  \centering
  \includegraphics[scale=0.36]{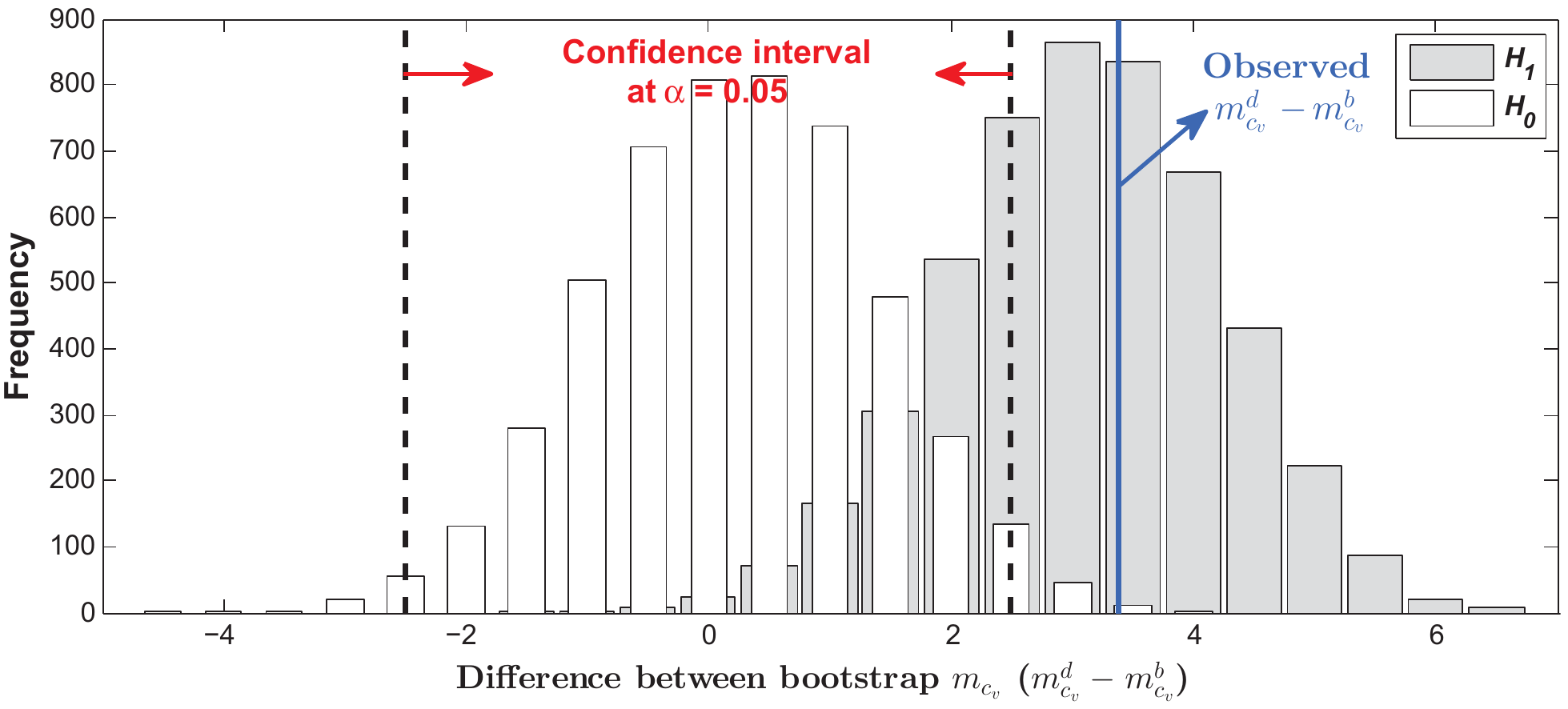}\\
 \caption{Our data support anecdote success stories for $\venue_P$.}
  \label{fig:boot_dist}
\end{figure}

As mentioned in the introduction there are various anecdote stories supporting the effectiveness of promotions through LBSNs such as the one for $\venue_P$.
At this part of our study we want to examine what our data imply for this specific venue and to verify whether our data and analysis are able to recover known ground truth.
$\venue_P$ publishes a special deal on the 37$^{th}$ day of the data collection, which lasts until the end of the collection period.
Therefore, we can only examine the short-term effectiveness.  
The standardized effect size observed is approximately $0.52$, while our bootstrap test indicates that this increase is statistically significant.
This is in complete agreement with reports about the specific venue \cite{Fsq-success-stories}.
Figure \ref{fig:boot_dist} further presents the bootstrap distribution of $\median_{\norcheckinst_\venue}^d-\median_{\norcheckinst_\venue}^b$ under $H_0$ and $H_1$.

\section{Models for Local Promotions}
\label{sec:models}

In this section we want to examine whether there are specific attributes that contribute to the success of a promotion.
For this we build models that can provide an educated decision on whether a special deal will ``succeed'' or not considering the short 
and the long term separately treating them as two different binary classification problems.  
Based on the bootstrap tests the positive class includes the offers that exhibit statistically significant increase in $\median_{\norcheckinst_\venue}^d$ ($\median_{\norcheckinst_\venue}^a$), while the negative class includes the special deals with a statistically significant decrease or a failure to reject the null hypothesis with a powerful test ($\pi \ge 0.8$).
We begin by extracting three different types of features.
Note that some of these features are specific to promotions, while others aim to capture other factors that can affect the popularity of a venue in general (e.g., neighborhood urban form etc.).
We then evaluate the predictive power of each individual feature using a simple unsupervised learning classifier.  
We further build a supervised learning classifier to predict the effect of a special deal using the extracted features.

\subsection{Feature extraction}
\label{sec:features}

\subsubsection{Venue-based features ($\mathcal{F}_{\venue}$): }
The set $\mathcal{F}_{\venue}$ includes features related with the properties of the venue publishing the special deal.
The intuition behind extracting such type of features lays on the fact that the effectiveness of the special offer can be connected to the characteristics of the venue itself.
For instance, a special deal might not help at all a really unpopular venue but it might be a great boost for a venue with medium levels of popularity.

{\bf Venue type: } This is the top-level type $\type(\venue)$ of venue {\venue}.
Table \ref{tab:feature_venue_category} depicts the fraction of special deals offered from different types of venues that are associated with a statistically significant increase in the daily number of check-ins ${\norcheckinst}$; i.e., the conditional probability $P(I | \type(\venue))$.

\begin{table*}[ht]
\small
\begin{center}
\tbl{Probability for the positive class conditioned on the type of the venue.  \label{tab:feature_venue_category}}\vspace{-0.03in}{
\begin{tabular}{|c|l|r|r|r|r|r|r|r|r|r|}
\hline
 \multicolumn{2}{|c|}{Category} & Nightlife & Food & Shops & Arts & College & Outdoors & Travel & Residence & Professional \\
\hline
 \% Positive & {\em short-term} & 62.07\% & 57.74\% & 42.90\% & 52.87\% & 56.25\% & 58.33\% & 66.84\% & 54.54\% & 61.86\%\\
 \hhline{~----------}
 class & {\em long-term} & 50.00\% & 41.51\% & 28.22\% & 43.75\% & 37.04\% & 25.00\% & 53.80\% & 14.29\% & 39.68\%\\
\hline
\end{tabular}}
\end{center}
\end{table*}

{\bf Popularity: } For the venue popularity we use two separate features;
(i) the mean number of check-ins per day at the venue for the period before the special offer starts, $\median_{\norcheckinst_\venue}^b$ and (ii) the cumulative number of check-ins in {\venue} just before the beginning of the special offer, ${\checkinst}_{\venue}[t_{s-1}]$.

{\bf Loyalty: } We define the loyalty $\loyalty$ of users in venue {\venue} as:

\begin{equation}
\loyalty_\venue[t_{s-1}] = \frac{{\checkinst}_{\venue}[t_{s-1}]}{{\uniquevt}_{\venue}[t_{s-1}]}
\label{eq:stickiness}
\end{equation}

\noindent where ${\uniquevt}_{\venue}[t_{s-1}]$ is the accumulated number of unique users that have checked-in to venue {\venue} at time $t_{s-1}$.
At a high-level $\loyalty$ indicates the average return (check-in) rate of users in \venue.

{\bf Likes: } Foursquare allows users to like or dislike a venue.
We will use the accumulated number of likes $\likes_{\venue}[t_{s-1}]$ a venue has received (at time $t_{s-1}$) as a feature for our classifiers.

{\bf Tips: } Foursquare allows users to leave short reviews for the venues.  We use the total number of such reviews (tips in Foursquare's terminology) $\tips_{\venue}[t_{s-1}]$ for venue {\venue} up to time $t_{s-1}$ as a feature for our classifiers.

\subsubsection{Promotion-based features ($\mathcal{F}_p$): }
The set $\mathcal{F}_p$ includes features related to the details of the special offer(s) that exist during the promotion period.
The details of the deal(s) might be important on whether the promotion will succeed or not.
For instance, a short-lived offer might have no impact because people did not have a chance to learn about it.

{\bf Duration: } The duration $\duration$ is the promotion period length. Intuitively, a longer duration allows users to learn and ``spread the word'' about the promotion, which consequently will attract more customers to check-in to the venue.

{\bf Type: } There are 7 types of special deals that can be offered from a Foursquare venue during the promotion period.  Each type provides different kind of benefits but has also different unlocking constrains. Table \ref{tab:feature_special_types} shows the probability distribution of the positive class conditioned on the different types of special offers that are part of the promotion.

\begin{table*}[htbp]
\small
\begin{center}
\tbl{Probability distribution of the positive class conditioned on the different types of special offers. \label{tab:feature_special_types}}\vspace{-0.03in}{
\begin{tabular}{|c|l|c|c|c|c|c|c|c|c|}
\hline
 \multicolumn{2}{|c|}{Type} & Newbie & Flash & Frequency & Friends & Mayor & Loyalty & Swarm & Multi-type \\
\hline
 \% Positive & {\em short-term} & 62.24\% & 60.00\% & 45.56\% & 84.62\% & 67.74\% & 50.50\% & 57.14\% & 60.60\% \\ 
 \hhline{~---------}
 class & {\em long-term} & 59.32\% & 62.50\% & 30.07\% & 43.75\% & 54.84\% & 50.00\% & 0.00\% & 44.23\% \\
\hline
\end{tabular}}
\end{center}
\end{table*}
If a venue publishes two (or more) different types of deals we refer to this as ``Multi-type'' offer.
In order to be able to easily distinguish between different combinations of offers in this ``Multi-type'' deals, we encode this categorical feature in a binary vector $\specialTypeVector \in \{0,1\}^7$, where each element represents a special type.
``Multi-type'' promotions will have multiple non-zero elements.

{\bf Count: } Count $\specialCount$ is the average number of special deals per day associated with a promotion period.  
$\specialCount$ captures how frequently a venue published specials during a specific promotion period.
Note that $\specialTypeVector$ is a binary vector and hence, if a venue is offering two deals of the same type this can only be captured through $\specialCount$.

\subsubsection{Geographical features ($\mathcal{F}_g$): }
The effectiveness of a promotion can be also related to the urban business environment in the proximity of the venue.
The latter can be captured through the spatial distribution of venues.
For example, an isolated restaurant might not benefit from a special deal promotion, simply because people do not explore the specific area for other attractions.
For our analysis, we consider the neighborhood $\neighborhood(\venue,r)$ of a venue {\venue} to be the set of venues within distance $r$ miles from {\venue} (we use $r=0.5$).

{\bf Density: } We denote the number of neighboring venues around {\venue} as the density $\neighborhooddensity_{\venue}$ of $\neighborhood(\venue,r)$.
Hence,

\begin{equation}
\neighborhooddensity_{\venue} = |\neighborhood(\venue,r)|
\label{eq:ndensity}
\end{equation}

{\bf Area popularity: }
The density $\neighborhooddensity_{\venue}$ captures a static aspect of {\venue}'s neighborhood.
To capture the dynamic aspect of the overall popularity of the area, we extract the total number of check-ins observed in the neighborhood at time $t_{s-1}$:

\begin{equation}
\areapopularity_{\venue} = \sum_{{\venue}'\in \neighborhood(\venue,r)} {{\checkinst}_{\venue'}[t_{s-1}]}
\label{eq:area_pop}
\end{equation}

Intuitively, a more popular area could imply higher likelihood for Foursquare users and potential customers to learn about the promotion and be influenced to visit the venue.

{\bf Competitiveness: }
A venue {\venue} of type $\type(\venue)$, will compete for customers only with neighboring venues of the same type.
Hence, we calculate the proportion of neighboring venues that belong to the same type $\type({\venue})$:

\begin{equation}
\competitiveness_{\venue} = \frac{|\venue' \in \neighborhood(\venue,r) \land \type(\venue')=\type(\venue)|}{\neighborhooddensity_{\venue}}
\label{eq:compete}
\end{equation}

{\bf Neighborhood entropy: } Apart from the business density of the area around {\venue}, the diversity of the local venues might be important as well.  To capture diversity we typically rely on the concept of information entropy.
In our setting we calculate the entropy of the distribution of the venue types in $\neighborhood(\venue,r)$.
With $f_{\type}$ being the fraction of venues in  $\neighborhood(\venue,r)$ of type {\type} the entropy of the neighborhood around {\venue} is:

\begin{equation}
\entropy_{\venue} = - \sum_{\type \in \types} f_{\type}\cdot log(f_{\type})   
\label{eq:entropy}
\end{equation}

\noindent where, {\types} is the set of all (top-level) venue types.

\subsection{Predictive power of individual features}
\label{sec:unsupervised}

We now examine the predictive ability of each of the numerical features described above in isolation.  
We will compare descriptive statistics of the distribution of each feature (in particular the median)
for the two classes.
We will then compute the ROC curve for each feature considering a simple, threshold-based, unsupervised classification system.

{\bf Mann-Whitney U test for each feature's median: }
A specific numerical feature $X$ can be thought of as being strongly discriminative for a classification problem, if the distributions of $X$ for the positive and negative instances are ``significantly'' different.
To that end we examine the sample median of these distributions by performing
the two-sided Mann-Whitney U test for the median values in the positive and negative classes for each of the features.
The $p$-values of these tests are presented in Table \ref{tab:feature_predictive_power_AUC}.

{\bf ROC curves for individual features: }
We now compute the ROC curve for each feature based on a simple unsupervised classifier.
The latter considers each feature $X$ in isolation and sets a threshold value for $X$ that is used to decide the class of every instance in our dataset.
For each value of this threshold we obtain a true-positive and false-positive rate.
We further calculate the area under the ROC curve (AUC).
Interestingly, there is a connection between the Mann-Whitney U test and the AUC given by \cite{cortes03}:

\begin{equation}
AUC = \frac{U}{n_p \cdot n_n}
\label{eq:auc_u-test}
\end{equation}

\noindent where $U$ is the value of the Mann-Whitney U test statistic, $n_p$ is the number of positive instances and $n_n$ is the number of negative instances.
Table \ref{tab:feature_predictive_power_AUC} presents the values for AUC.
As we observe while there are some features that deliver a good performance (e.g., $\median_{\norcheckinst_\venue}^b$ and $\specialCount$) most of the features give a performance close to the random baseline of 0.5.
Hence, each feature individually does not appear to be a good predictor for the effect of special offers through LBSNs.
However, in the following section we will examine a supervised learning approach utilizing combinations of the features.

\begin{table}[h]
\small
\begin{center}
\tbl{While the median of the features for the two classes are significantly different, the actual distribution appear to not be discriminative (low AUC). 
\label{tab:feature_predictive_power_AUC}}\vspace{-0.03in}{
\begin{tabular}{|cc||r|r|r|r|}
\hline
 \multicolumn{2}{|c||}{\multirow{2}{*}{Features}} & \multicolumn{2}{c|}{short-term} & \multicolumn{2}{c|}{long-term} \\
 \hhline{~~----}
 & & AUC & $p$-value & AUC & $p$-value \\
\hline
 \multicolumn{1}{|c|}{\multirow{6}{*}{$\mathcal{F}_{\venue}$}} & ${\checkinst}_{\venue}[t_{s-1}]$ & 0.537 & $10^{-6}$ & 0.519 & $0.047$ \\
\hhline{~-----}
 \multicolumn{1}{|c|}{} & $\median_{\norcheckinst_{\venue}}^b$ & {\bf 0.799} & $0$ & {\bf 0.702} & $0$\\
\hhline{~-----}
 \multicolumn{1}{|c|}{} & $\loyalty_{\venue}[t_{s-1}]$  & 0.526 &  $10^{-4}$ & 0.535 & $10^{-4}$\\
\hhline{~-----}
 \multicolumn{1}{|c|}{} & $\likes_{\venue}[t_{s-1}]$ & 0.537 &  $10^{-9}$  & 0.557 & $0$\\
\hhline{~-----}
 \multicolumn{1}{|c|}{} & $\tips_{\venue}[t_{s-1}]$  & 0.510 & 0.178 & 0.546 & $10^{-7}$ \\
\hline
 \multicolumn{1}{|c|}{\multirow{2}{*}{$\mathcal{F}_p$}} & $\duration$ & 0.539 & $10^{-7}$ & 0.520 & $0$ \\
\hhline{~-----}
 \multicolumn{1}{|c|}{} & $\specialCount$ & {\bf 0.617} & $0$ & {\bf 0.609} & $0$ \\
\hline
 \multicolumn{1}{|c|}{\multirow{4}{*}{$\mathcal{F}_g$}} & $\neighborhooddensity_{\venue}$ & 0.551 & $0$ & 0.551 & $10^{-8}$ \\
\hhline{~-----}
 \multicolumn{1}{|c|}{} & $\areapopularity_{\venue}$ & 0.558 & $0$ & 0.558 & $10^{-9}$ \\
\hhline{~-----}
 \multicolumn{1}{|c|}{} & $\competitiveness_{\venue}$ & 0.565 & $0$ & 0.557 & $10^{-9}$ \\
\hhline{~-----}
 \multicolumn{1}{|c|}{} & $\entropy_{\venue}$ & 0.559 & $0$ & 0.574 & $0$ \\
\hline
\end{tabular}}
\end{center}
\end{table}

\subsection{Supervised learning classifiers}
\label{sec:supervised}

In this section we turn our attention to supervised learning models and we combine the extracted features to improve the classification performance achieved by each one of them individually.
We evaluate various combinations of the three types of features, while our performance metrics include accuracy, F-measure and AUC.
Furthermore, we examine two different models, a linear one (i.e., logistic regression) and a more complex based on ensemble learning (i.e., random forest).

We begin by evaluating our models through 10-fold cross validation on our labeled promotion dataset.
The results for the different combinations of features and for the different classifiers are shown in Table \ref{tab:classifier_treatment}.
As the results indicate, even when we use simple linear models the performance is significantly improved compared to unsupervised models.
It is also interesting to note that the most important type of features appears to be the venue-based features $\mathcal{F}_{\venue}$.
The promotion-based as well as the geographic features while improving the classification performance when added, do not provide very large improvements.

The above models were built and evaluated on the data points identified through the bootstrap statistical tests in an effort to keep the false positives/negatives of the labels low.
However, while this is important for building a robust model, in a real-world application the model will need to output predictions for cases that might not provide statistically significant results a posteriori.
After all, a venue owner is interested in what he observes, and not whether this was a false positive/negative (i.e., an increase/decrease that happened by chance).
Hence, we test the performance of our models on the data points in the promotion group for which we were not able to identify a statistically significant change ($\alpha=0.05$) in the average number of check-ins per day.
A positive observed value of $d$ corresponds to the positive class.
Note that we do not use these points for training.
This resembles an out-of-sample evaluation of our models, testing their generalizability to less robust observations.
Our results are presented in Table \ref{tab:experiment3}.
As we can see, while as one might have expected the performance is degraded compared to the cross-validation setting, it is still good.

\begin{table*}[ht]
\small
\begin{center}
\tbl{Using supervised learning models improves the performance over unsupervised learning methods. \label{tab:classifier_treatment}}\vspace{-0.03in}{
\begin{tabular}{|c|c||c|c|c||c|c|c|}
\hline
 \multirow{2}{*}{Algorithm} & \multirow{2}{*}{Feature} & \multicolumn{3}{c||}{short-term} & \multicolumn{3}{c|}{long-term} \\
 \hhline{~~------}
 &  & Accuracy & F-measure & AUC & Accuracy & F-measure & AUC\\
\hline

         & $\mathcal{F}_p$ & 0.582 & 0.474 & 0.583 & 0.684 & 0.139 & 0.642 \\
         & $\mathcal{F}_{\venue}$ & 0.831 & 0.836 & 0.882 & 0.826 & 0.68 & 0.876 \\
Logistic & $\mathcal{F}_g$ & 0.569 & 0.532 & 0.579 & 0.686 & 0.029 & 0.582 \\
Regression& $\mathcal{F}_p$$\cup$$\mathcal{F}_{\venue}$ & 0.833 & 0.835 & 0.885 & 0.831 & 0.697 & 0.876 \\
         & $\mathcal{F}_p$$\cup$$\mathcal{F}_g$ & 0.588 & 0.52 & 0.618 & 0.684 & 0.128 & 0.641 \\
         & $\mathcal{F}_{\venue}$$\cup$$\mathcal{F}_g$ & 0.83 & 0.835 & 0.882 & 0.827 & 0.687 & 0.876 \\
         & $\mathcal{F}_p$$\cup$$\mathcal{F}_{\venue}$$\cup$$\mathcal{F}_g$ & 0.834 & 0.836 & 0.885 & 0.833 & 0.704 & 0.876 \\
\hline

         & $\mathcal{F}_p$ & 0.681 & 0.672 & 0.76 & 0.685 & 0.349 & 0.702\\
         & $\mathcal{F}_{\venue}$ & 0.856 & 0.846 & 0.931 & 0.86 & 0.761 & 0.9\\
  Random & $\mathcal{F}_g$ & 0.559 & 0.523 & 0.578 & 0.646 & 0.285 & 0.576\\
  Forest & $\mathcal{F}_p$$\cup$$\mathcal{F}_{\venue}$ & 0.87 & 0.862 & 0.943 & 0.868 & 0.777 & 0.909\\
         & $\mathcal{F}_p$$\cup$$\mathcal{F}_g$ & 0.666 & 0.652 & 0.74 & 0.685 & 0.396 & 0.697\\
         & $\mathcal{F}_{\venue}$$\cup$$\mathcal{F}_g$ & 0.856 & 0.846 & 0.934 & 0.862 & 0.765 & 0.904\\
         & $\mathcal{F}_p$$\cup$$\mathcal{F}_{\venue}$$\cup$$\mathcal{F}_g$ & 0.87 & 0.861 & 0.94 & 0.863 & 0.765 & 0.91\\
\hline
\end{tabular}}
\end{center}
\end{table*}

\begin{table*}[ht]
\small
\begin{center}
\tbl{Our supervised models deliver good performance on out-of-sample evaluation on the less robust observations. \label{tab:experiment3}}\vspace{-0.03in}{
\begin{tabular}{|c|c||c|c|c||c|c|c|}
\hline
 \multirow{2}{*}{Algorithm} & \multirow{2}{*}{Feature} & \multicolumn{3}{c||}{short-term} & \multicolumn{3}{c|}{long-term} \\
 \hhline{~~------}
 &  & Accuracy & F-measure & AUC & Accuracy & F-measure & AUC\\
\hline

         & $\mathcal{F}_p$ & 0.484 & 0.353 & 0.486 & 0.532 & 0.21 & 0.543 \\
         & $\mathcal{F}_{\venue}$ & 0.625 & 0.696 & 0.678 & 0.589 & 0.436 & 0.659 \\
 Logistic& $\mathcal{F}_g$ & 0.497 & 0.442 & 0.5 & 0.518 & 0.011 & 0.527 \\
Regression& $\mathcal{F}_p$$\cup$$\mathcal{F}_{\venue}$ & 0.628 & 0.683 & 0.654 & 0.596 & 0.509 & 0.651 \\
         & $\mathcal{F}_p$$\cup$$\mathcal{F}_g$ & 0.491 & 0.389 & 0.494 & 0.524 & 0.133 & 0.553 \\
         & $\mathcal{F}_{\venue}$$\cup$$\mathcal{F}_g$ & 0.625 & 0.695 & 0.677 & 0.592 & 0.459 & 0.657 \\
         & $\mathcal{F}_p$$\cup$$\mathcal{F}_{\venue}$$\cup$$\mathcal{F}_g$ & 0.627 & 0.683 & 0.656 & 0.6 & 0.521 & 0.653 \\
\hline

         & $\mathcal{F}_p$ & 0.532 & 0.552 & 0.54 & 0.52 & 0.338 & 0.56\\
         & $\mathcal{F}_{\venue}$ & 0.641 & 0.681 & 0.676 & 0.608 & 0.628 & 0.628\\
  Random & $\mathcal{F}_g$ & 0.503 & 0.469 & 0.508 & 0.522 & 0.283 & 0.513\\
  Forest & $\mathcal{F}_p$$\cup$$\mathcal{F}_{\venue}$ & 0.639 & 0.681 & 0.676 & 0.611 & 0.63 & 0.631\\
         & $\mathcal{F}_p$$\cup$$\mathcal{F}_g$ & 0.525 & 0.541 & 0.539 & 0.526 & 0.356 & 0.547\\
         & $\mathcal{F}_{\venue}$$\cup$$\mathcal{F}_g$ & 0.643 & 0.682 & 0.678 & 0.61 & 0.627 & 0.63\\
         & $\mathcal{F}_p$$\cup$$\mathcal{F}_{\venue}$$\cup$$\mathcal{F}_g$ & 0.643 & 0.682 & 0.677 & 0.612 & 0.628 & 0.631\\
\hline
\end{tabular}}
\end{center}
\end{table*}

Finally we focus on the results from logistic regression, which has a genuine probabilistic interpretation.
In particular, the accuracy performance when using the set of features $\mathcal{F}_{\venue}$$\cup$$\mathcal{F}_g$  and $\mathcal{F}_p$$\cup$$\mathcal{F}_{\venue}$$\cup$$\mathcal{F}_g$ is very similar.
We compute the actual outcome of the model, i.e., before applying the classification threshold, which is the probability of observing an increase in the mean daily check-ins of the corresponding venue.
Hence, the outcome of the two models provide the probabilities $P(I|\mathcal{F}_{\venue},\mathcal{F}_g)$ and $P(I|\mathcal{F}_{\venue},\mathcal{F}_g,\mathcal{F}_p)$ respectively.
We calculate the difference between these probabilities for all the corresponding cases in Tables \ref{tab:classifier_treatment} and \ref{tab:experiment3}.
Table \ref{tab:predicted probability_distance} presents the root mean square differences, which is small for all the scenarios.
Since features $\mathcal{F}_{\venue}$ and $\mathcal{F}_g$ capture various (environmental) externalities, while the set $\mathcal{F}_p$ captures attributes related with the promotion itself, these results further support our findings from our statistical analysis.  
Of course these features do not capture all the externalities, and thus the actual probabilities might differ, even though the classification outcome is very accurate.

\begin{table}[h]
\small
\begin{center}
\vspace{-0.03in}
\tbl{The root mean square distance of the logistic regression output for the features $\mathcal{F}_{\venue}$$\cup$$\mathcal{F}_g$  and $\mathcal{F}_p$$\cup$$\mathcal{F}_{\venue}$$\cup$$\mathcal{F}_g$ further supports our statistical analysis.
\label{tab:predicted probability_distance}}\vspace{-0.1in}{
\begin{tabular}{|c|c|c|c|}
\hline
 \multicolumn{2}{|c|}{Cross-validation} & \multicolumn{2}{c|}{Out-of-sample} \\
\hline
 short-term & long-term & short-term & long-term \\
\hline
 0.081 & 0.067 & 0.072 & 0.074 \\
\hline
\end{tabular}}
\end{center}
\end{table}

\section{Related Work}
\label{sec:related}

{\bf Effects of Promotions: }
There are studies in the management science that examine the impact of promotions 
on marketing.
For example, \cite{blattberg1995promotions} found that temporary discounting substantially increases short term brand sales.  However, its long term effects tend to be much weaker.
This pattern was further quantified by \cite{pauwels2002long} who found that the significant short time promotion effects on customer purchases  
die out in subsequent weeks or months.
Furthermore, Srinivasan \textit{et al.} \cite{srinivasan2004promotions} quantified the price promotion impact on two targeted variables, namely, revenues and total profits, by using vector autoregressive modeling.
The authors found that the price promotion has a positive impact on manufacture revenues, but for retailers it depends on multiple factors such as brand and promotion frequency.
Finally, \cite{kopalle1999dynamic} proposed a descriptive dynamic model 
which suggests that the higher-share brands tend to over-promote (i.e., offer promotions very frequently), while the lower-share brands do not promote frequently enough.

{\bf Online Deals and Advertising: }
Online promotions 
have gained a lot of attention in recent literature.
Such promotions have been a popular strategy for local merchants to increase revenues and/or raise the awareness of potential customers.
A detailed business model analysis on Groupon was first presented by \cite{arabshahi2010undressing}, while in \cite{dholakia2010effective} the authors surveyed businesses that provide Groupon deals to determine their satisfaction.
Edelman {\em et al.} \cite{edelman2011groupon} considered the benefits and drawbacks from a merchant's point of view on using Groupon and provided a model that captures the interplay between advertising and price discrimination effects and the potential benefits to merchants.
Finally, Byers {\em et al.} \cite{byers2012daily} designed a predictive model for the Groupon deal size by combining features of the offer with information drawn from social media.  
They further examined the effect of Groupon deals on Yelp rating scores.  

Tangential to our work is also literature on web advertising and its efficiency.  
In this space, Fulgoni {\em et al.} \cite{fulgoni2008online} present data for the positive impact of online display advertising on search lift and sale lift, while Goldfarb {\em et al.} \cite{goldfarb2011online} further examined the effect of different properties of display advertising on its success through traditional user surveys.
Papadimitriou {\em et al.} \cite{papadimitriou2011display} study the impact of online display advertising on user search behavior using a controlled experiment.

{\bf Mobile Marketing and Social Media: }
Mobile marketing serves as a promising strategy for retail businesses to attract, maintain and enhance the connection with their customers.
Sliwinski \cite{sliwinski2002spatial} built a prototype application that utilizes customer spatial point pattern analysis to target potential new customers.
Furthermore, Banerjee \textit{et al.} \cite{banerjee2008mobile} studied the effectiveness of mobile advertising.
Their findings indicate that the actual location of the participant as well as the context of that location, significantly influence the potential effectiveness of these advertising strategies.
Recently, there have also been efforts to quantify through models \cite{baccelli11} the financial value of location data, which are in the center of mobile marketing operations.

In another direction, 
location-based social media have gained a lot of attention.
Data collected from such platforms can drive novel business analysis.
Qu and Zhang \cite{qu2013trade} proposed a framework that extends traditional trade area analysis and incorporates location data of mobile users.
As another example, Karamshuk {\em et al.} \cite{karamshuk2013geo} proposed a machine learning framework to predict the optimal placement for retail stores, where they extracted two types of features from a Foursquare check-in dataset.
Furthermore, these platforms can serve as mobile ``yellow pages'' with business reviews that can influence customer choices.
For example,  Luca \cite{luca2011reviews} has identified a causal impact of Yelp ratings on restaurant demand using the regression discontinuity framework.

\section{Discussion and Limitations} 
\label{sec:scope}

We would like to reiterate that this study should not be seen as a study on Foursquare per se.
Our work is focused on the mechanism of promotions through location-based social media.
Our results suggest that the benefits from local promotions through LBSNs are more limited than what anecdote stories suggest.
However, we acknowledge again that the time-series of daily check-ins is only a proxy for the actual revenue generated.
Nevertheless, we believe that even if the specific check-ins do not lead to direct revenue, they increase the {\em visibility} of the venue, at least within the ecosystem of social media.  
Note here that even though the potential of special campaigns through geo-social media appears to be limited, even a small increase in the probabilities $P({\increaseevent}_d |  \specialevent, \externalitiesevent)$ and $P({\increaseevent}_a |  \specialevent, \externalitiesevent)$ as compared to $P({\increaseevent}_d | \externalitiesevent)$ and $P({\increaseevent}_a |  \externalitiesevent)$ respectively, can still be deemed as a successful advertising model, given that typical conversion rates of online advertisements can be as small as 1\% \cite{kazienko15}.
Our analysis can also shed light on possible tweaks of the way they are offered.
For example, recent literature \cite{cramer11} has brought onto surface possible reasons that lead people to check-in to a location long after they arrive.
This means that these users might have not even used the social application to explore the area they are in and thus, they have not been aware at all about a special deal that was in the vicinity.
Therefore, more active communication channels for these campaigns might be required (e.g., geo-fenced push notifications).
The way that a promotion is redeemed might also play a role.
For example, some deals require users to have an American Express card. 
Furthermore, venues might combine their online promotions with other offline campaigns that can further improve the effectiveness of both advertising means.
Unfortunately, our analysis cannot account for this due to the lack of appropriate information.


From a technical point of view we have used bootstrap techniques for our hypothesis tests 
in order to avoid strong assumptions of standardized tests.
Nevertheless, bootstrap relies on the assumption that the obtained sample is representative of the population.  
In our case the {\em representativeness} of the sample might be challenged by its possibly small size (e.g., for promotions periods that last only one week).
Furthermore, the interpolation performed on the raw time-series might have added noise on the empirical bootstrap distribution obtained.
However, we expect that both of the distributions under the null and alternative hypotheses to have been affected in a similar manner, {\bf if at all}, and hence their relative positions to not have been affected.
In addition, while we have performed block bootstrap resampling with block size of 2 in order to account for dependencies between check-ins of consecutive days, 
the dependencies might be more complicated.  
Finally, the quality of the reference group can be significantly affected by the venues that have been created on the social media platform.
While we have accounted for this, identifying {\em spam} venues is beyond the scope of this work.
Moreover, if in addition to the type and location of a venue there are other observed or unobserved confounding factors that affect the decision to offer a promotion, our  reference groups might not be able to effectively account for self-selection biases.

\section{Conclusions and Future Work}
\label{sec:conclusions}

We study the effectiveness of special deals that local establishments can offer through LBSNs.
We collect and analyze a large dataset from Foursquare using randomization and statistical bootstrap.
We find that promotions through LBSNs do not alter the probability of observing an increase in the daily check-ins to a venue, while the underlying standardized effect size changes only slightly.
We also model the effectiveness of such offers by extracting three different types of features and building classifiers that can provide us with an educated decision with regards to the success of these promotions.
In the future, we opt to incorporate into our analysis the lower level categories for the locales as well as examine alternative evaluation metrics (e.g., number of unique users).
Finally, we plan to explore ways to study a recently introduced mechanism for advertisements through LBSNs \cite{Fsq-ads}, which appears to be effective based again on anecdotes.  

\fontsize{9.5pt}{10.5pt}
\selectfont
\bibliographystyle{aaai}
\bibliography{main}

\end{document}